\documentclass[11pt,onecolumn,draftcls]{IEEEtran}
\usepackage[T1]{fontenc}
\usepackage{amsmath,amssymb}
\usepackage{graphicx}
\usepackage{epsfig}
\usepackage{cite}
\usepackage{setspace}
\usepackage{color}
\usepackage{enumerate}
\usepackage{multirow}

\title{Coalitional Games for Transmitter Cooperation in MIMO Multiple Access Channels}
\author{Srinivas Yerramalli, Rahul Jain and Urbashi Mitra \\
    srinivas.yerramalli, rahul.jain, ubli@usc.edu \\
    University of Southern California, Los Angeles, CA
    \thanks{This research has been funded in part by the following grants and organizations: NSF CCF-0917343, NSF IIS 0917410 and CAREER CNS-0954116. Some of the results in this paper have appeared in ISIT 2011\cite{Srinivas11ISIT} and SPCOM 2012 \cite{Srinivas12SPCOM}. The authors would like to thank Prof. Upamanyu Madhow for suggestions in the consideration of time sharing with SIC.}}

\newtheorem{thm}{Theorem}
\newtheorem{mydef}{Definition}
\newtheorem{lemma}{Lemma}

\newtheorem{prop}{Proposition}

\newcommand{\tQ}{\tilde Q}
\newcommand{\hQ}{\hat Q}
\newcommand{\tB}{\tilde B}
\newcommand{\hB}{\hat B}
\newcommand{\uh}{{\underline h}}

\begin{document}
\maketitle
\thispagestyle{empty}

\vspace{-0.5in}
\begin{abstract}
Cooperation between nodes sharing a wireless channel is becoming increasingly necessary to achieve performance goals in a wireless network. The problem of determining the feasibility and stability of cooperation between rational nodes in a wireless network is of great importance in understanding cooperative behavior. This paper addresses the stability of the grand coalition of transmitters signaling over a multiple access channel using the framework of cooperative game theory. The external interference experienced by each TX is represented accurately by modeling the cooperation game between the TXs in \emph{partition form}. Single user decoding and successive interference cancelling strategies are examined at the receiver. In the absence of coordination costs, the grand coalition is shown to be \emph{sum-rate optimal} for both strategies. Transmitter cooperation is \emph{stable}, if and only if the core of the game (the set of all divisions of grand coalition utility such that no coalition deviates) is nonempty. Determining the stability of cooperation is a co-NP-complete problem in general. For a single user decoding receiver, transmitter cooperation is shown to be \emph{stable} at both high and low SNRs, while for an interference cancelling receiver with a fixed decoding order, cooperation is stable only at low SNRs and unstable at high SNR. When time sharing is allowed between decoding orders, it is shown using an approximate lower bound to the utility function that TX cooperation is also stable at high SNRs. Thus, this paper demonstrates that ideal zero cost TX cooperation over a MAC is stable and improves achievable rates for each individual user.

\end{abstract}

\section{Introduction}
\label{sec:intro}
Next generation wireless networks are being designed to operate in a complex and dynamic environment in which nodes interact and cooperate to improve network throughput (see \cite{Saad09,Sendonaris03,Larsson09,survey11,Samson09,Saad09-TWC} and the references therein). As nodes signaling via wireless share resources due to the broadcast nature of the medium, cooperation between such nodes has emerged as a key strategy for improving performance \cite{Jindal04}. In typical cooperative scenarios, it is inherently assumed that all nodes are controlled centrally and hence cooperation can be enforced. However, the emergence of heterogenous networks without a unified central controller challenges this assumption. In such scenarios, it is reasonable to assume that a rational node would willingly cooperate only if cooperation improves its own utility. The problem of determining which nodes in a network would cooperate in a stable fashion and how the benefits of such cooperation would be shared has thus become important, especially for these future heterogenous networks. If the properties of cooperation are well understood in elementary networks such as the multiple access channel (MAC), the broadcast channel (BC), the interference channel (IC) \textit{etc.,} larger networks can be viewed as a composition of several elementary networks and cooperative behavior can then be examined to draw useful insights for system design. Towards this goal, this paper addresses the problem of TX cooperation in a MAC.

\begin{figure}
\centering{
    \includegraphics[scale=0.5]{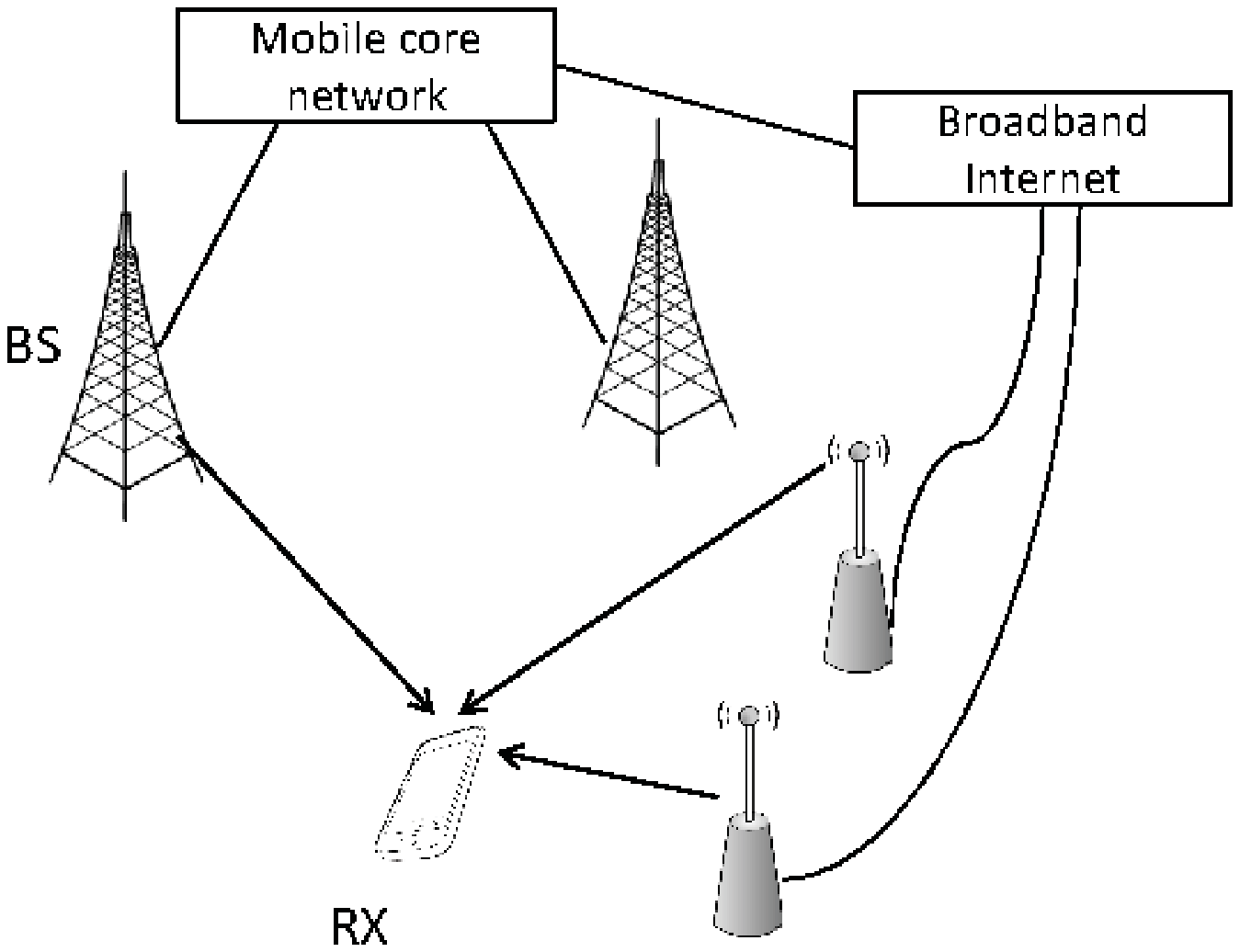}
    \caption{Cooperative signaling between base station and femto cell network for joint transmission to UE}
\label{fig:network}}
\end{figure}

Consider a example scenario as shown in Fig. \ref{fig:network} in which several (service providers) base stations and femtocells (TXs) operating in the same frequency band are in the range of a mobile (RX). The base stations and the femtocells can be connected via a backbone wired network which can be used to share channel state and codebook information. However, each TX may be owned by a different operator and each operator would like to provide the largest rate to the mobile to increase his revenue. A \emph{rational} TX with an intent to maximize revenue, could cooperate with other TXs in the range of the mobile by joint encoding and transmission to improve data rates of the mobile (total sum rate) and improve the data rates for each TX. In such a scenario, the TX cooperation game studied in this paper, can be used to determine the optimal coalition structure and the individual rates that each TX gets to transmit.

\begin{figure}
\centering{
    \includegraphics[scale=0.6]{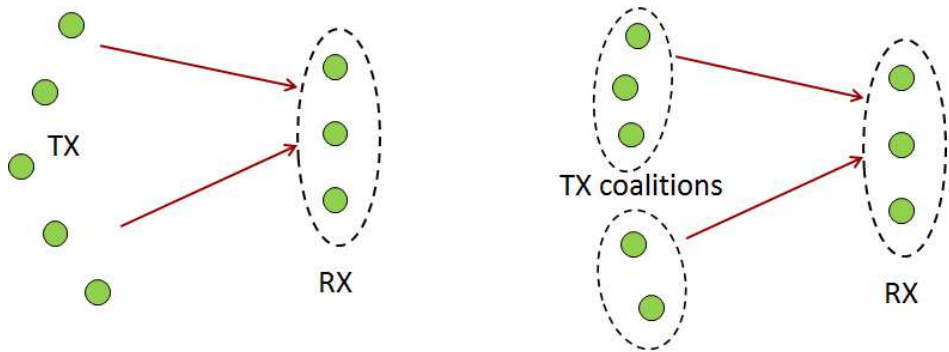}
\caption{Transmitters cooperate to form coalitions. All transmitters in a coalition fully cooperate with each other}
\label{fig:coalitions}}
\vspace{-0.25in}
\end{figure}

\emph{Non-cooperative} games between TXs in a MAC has been analyzed for various scenarios such as fading multiple-input single-output (MISO) channels in \cite{Lai08}, for fading MIMO channels in \cite{Belmega10} and for incomplete channel state information in \cite{He10}. To enhance the sum rate, rational users (TXs here) can form coalitions and cooperate by signaling jointly to the RX (see Fig. \ref{fig:coalitions}). In the absence of cooperation costs, there is an incentive to form the \emph{grand coalition} (GC), the coalitions of all users, and signal jointly such that no data streams are decoded in the presence of interference. A larger coalition implies that the benefits of cooperation have to be shared among many users and the GC is stable only when no coalition of users has an incentive to break away and form a smaller coalition. The key question is to determine whether the gains of the GC are sufficient to share the payoff amongst all its members such that no coalition of TXs has an incentive to defect, thus ensuring stable cooperation. This is equivalent to checking the nonemptiness of the \emph{core}, the feasible set of a linear program which describes the demands of each coalition. A \emph{nonempty core} implies that the GC utility can be shared such that no coalition of TXs has an incentive to defect, ensuring its \emph{stability}. The issue of stability is one of the key questions addressed in this paper.

Several works in the literature have examined the nonemptiness of the core for TX cooperation in a Gaussian MAC under various assumptions, primarily using \emph{characteristic form games} (CFGs), in which the utility of a coalition is independent of the actions of members outside it. In \cite{La03}, it is assumed that TXs bargain for higher rates by threatening to jam transmissions. A key assumption in \cite{La03} is that the jamming users are not interested in transmitting data, in contrast to what we consider herein. In \cite{Mathur08}, the utility of each coalition of TXs is considered to be a minimum assured utility that can be obtained by assuming that all other TXs coordinate to jam the transmissions. This is a very conservative model of utility and it is shown in \cite{Mathur08} that the cooperation is unstable and exhibits oscillatory behavior in general. A packetized, slotted, version of the TX cooperation game was considered in \cite{Karamchandani} and cooperation was shown to be stable for certain scenarios. TX cooperation with non-zero cost of coalition formation in a slotted TDMA system was considered in \cite{Saad09-TWC}, and \cite{Madiman08} discusses the reinterpretation of several information theoretic results, including the MAC, from the point of view of cooperative game theory. As the interference generated by external TXs strongly influences the achievable rates in a Gaussian MAC, the CFG model considered in most previous works, does not accurately measure the interference as is needed to analyze TX cooperation in a classical Gaussian MAC. In \cite{Saad09}, the need for taking into account the actual interference that affects a coalition using PFGs \cite{Thrall63} has been stated and their potential for PFGs to provide a good framework for modeling and analysis of self-organizing next generation communication networks is discussed. \emph{The primary contribution of this paper is to study the problem of TX cooperation in a MAC under the framework of partition form game theory and show the stability of TX cooperation in several scenarios of interest.}
\begin{table}
\label{table:summary}
\begin{center}
\begin{tabular}{|c|c|c|c|}
\hline
Receiver & Regime &  Power Constraint Type & Core \\ \hline
\hline
\multirow{2}{*}{SUD} & high SNR & SP and PP & Nonempty \\
                     & low SNR  &  SP       & Nonempty \\  \hline
\multirow{2}{*}{SIC with fixed decoding order}  & high SNR & SP and PP & Empty in general \\
                                                & low SNR  & SP        & Nonempty \\ \hline
\multirow{2}{*}{SIC with time shared decoding orders}  & high SNR & SP and PP & Nonempty \\
                     & low SNR & SP & Nonempty \\ \hline
\end{tabular}
\end{center}
\caption{Table showing the summary of the nonemptiness of the core of the TX cooperation game for various scenarios. SP = sum power constraint and PP = per-antenna power constraint.}
\vspace{-0.35in}
\end{table}

Specifically, we consider a MAC with a single user decoding (SUD) RX which treats interference as noise and a successive interference cancellation (SIC) RX in which decoded signals are canceled out to reduce interference. The TX cooperation game is analyzed in several stages. The existence and uniqueness of a Nash equilibrium (NE) utility (we make the distinction between uniqueness of NE achievable strategies and NE utilities with the later being a weaker notion) for a \emph{non-cooperative} game between the TXs is first examined to determine the achievable rates for a given configuration of TXs. Next, TXs are assumed to cooperate to form larger coalitions and the change in utility from cooperation is characterized for both the cooperating and the external TXs. Using the utilities determined previously, we consider the various cores of a PFG \cite{Hafalir} (PFGs have several different definitions of the core, based on the expected behavior of external coalitions) and examine the the stability of cooperation by investigating their nonemptiness. Table \ref{table:summary} shows a summary of results for the various scenarios considered in this paper. For an SUD receiver, cooperation is stable at both high and low SNRs, while for a SIC receiver with a given decoding order, cooperation is stable only in the noise-limited regime and may be empty at high SNR. This can be attributed to the asymmetry between the users caused by a fixed decoding order. However, if time sharing between decoding orders is permitted, we show using a high SNR approximation to the utility function that cooperation is stable at high SNR as well\footnote{Note that while our results are true at high SNR for both sum power and per-antenna power constraints, the low SNR regime is characterized only for sum power constraints due to the lack of a suitable approximation of the capacity of a MIMO channel with per-antenna power constraints in this regime \cite{Vu11MIMO}.}. Thus, this paper demonstrates the role played by the choice of RX and the regime of operation in determining the stability of TX cooperation in a MIMO MAC.

The rest of the paper is organized as follows. Section \ref{sec:prelims} defines several useful concepts related to cooperative game theory and Section \ref{sec:signal_model} describes the signal model for a MIMO MAC. The stability of TX cooperation is analyzed for a SIC and SUD RX in Section \ref{sec:tx_cooperation_intf} and Section \ref{sec:tx_cooperation} respectively. Section \ref{sec:conclusions} concludes this paper.

\section{Cooperative Game theory - Preliminaries}
\label{sec:prelims}
We begin by reviewing several game theoretic preliminaries for cooperative games. Let $S \subseteq \mathcal{K}, \mathcal{K} = \left \{1,2,...,K \right \}$ denote an arbitrary coalition of TXs.
\begin{mydef}
A partition $T$ of $\mathcal{K}$ is defined as a set of coalitions $S_1,S_2,...,S_N$ such that $S_i \cap S_j = \phi,~ \forall i,j \in \left \{1,2,...,N \right \}, ~ i \neq j $ and $\cup_{i=1}^{N} S_i = \mathcal{K}$.
\end{mydef}
The set of all partitions of $\mathcal{K}$ is denoted by $\mathcal{T}$. The total number of partitions of a $K$-element set is called the \textit{Bell number} $B_K$ and increases exponentially in $K$.

In a cooperative game, players form coalitions and each coalition chooses an action from the set of actions \emph{jointly} available to it (which may be larger than the set of actions available individually to each of the players). By this choice of actions, each coalition $S$ in partition $T$ obtains a utility (value) denoted by $v(S;T)$. Games in which the utility of each coalition is dependent on the actions of other coalitions are called \emph{partition form games} (PFGs) while games in which utility is independent of external actions are called \textit{characteristic form games} (CFGs), \textit{i.e.,} $v(S;T) = v(S)$ is independent of the specific partition $T$.

\begin{mydef}
A coalitional game is called a \emph{transferable utility (TU) game}, if the cooperative gains achieved by a coalition can be arbitrarily divided among all members of the coalition.
\end{mydef}
In a TU game, that the payoff obtained by cooperation is given to the cooperating coalition to be divided among its members. In contrast, for a non-transferable utility game, cooperation results in payoffs to each individual member of the coalition directly and cannot be redistributed to the other members. In this paper, we consider a TU game and denote by $x_k$, the utility allocated to the $kth$ player.

\begin{mydef}
A PFG is said to be \emph{cohesive} if for any partition $T = \{ S_1,S_2,...,S_N \}$ of $\mathcal{K}$,
$ v(\mathcal{K}; \mathcal{K}) \geq \sum_{n=1}^{N} v(S_n;T) $.
\end{mydef}
For a cohesive game, the utility obtained by the GC is larger than the sum of utilities of each coalition under any other partition. In other words, the GC maximizes the sum utility among all configurations.

\begin{mydef}
A PFG is \emph{r-super-additive} if for any disjoint coalitions $ S_1, S_2, ..., S_r $ and any partition $\rho$ of $\mathcal{K} \setminus ( S_1 \cup S_2 \cup ... \cup S_r )$, we have $ v(S_1 \cup ... \cup S_r ; \left \{ S_1 \cup ... \cup  S_r \right \} \cup \rho ) \geq \sum_{t=1}^{r} v(S_t  ; \left \{ S_1, S_2,...,S_r \right \} \cup \rho ) $.
\label{e:r_super_additive}
\end{mydef}
Super-additivity implies that the when coalitions merge to form a larger coalition, the total utility of the larger coalition is greater than the sum of the utilities of its constituent coalitions. In other words, forming larger coalitions improves achievable utility. When $r=2$, the above definition reduces to the conventional definition of super-additivity in PFGs \cite{Hafalir}. For CFGs, the utility of the coalition is independent of the rest of the partition and hence $2$-super-additivity implies $r$-super-additivity and cohesiveness. This is however not true for PFGs due to externalities. However, it is clear that if a game is $r$-super-additive for all feasible values of $r$, then the game is cohesive and the GC has the maximum total utility.

\begin{mydef}
A PFG is said to have \emph{negative externalities} if for any mutually disjoint coalitions $ S_1, S_2$ and $ S_3 $ and any partition $\rho$ of $\mathcal{K} \setminus ( S_1 \cup S_2 \cup S_3)$, we have $ v(S_3; \{S_3, S_1 \cup S_2 \} \cup \rho ) \leq v (S_3 ; \{ S_1,S_2,S_3 \} \cup \rho ) $.
\label{e:negative_externalities}
\end{mydef}
\vspace{-0.2in}
A game with positive externalities is defined similarly with the inequality reversed. A game has negative (positive) externalities if a merger between two coalitions does not increase (decrease) the utility of all other coalitions. A game with mixed externalities exhibits both positive and negative externalities for different coalitions or for different realizations of the game parameters.

\subsection{Core and Stability of Cooperation in Partition Form Games}
The core of a cooperative game is the set of all divisions of utility such that no user or coalition of users has an incentive to deviate from a given configuration (usually the grand coalition). The core of a cooperative game can be represented as a linear program which describes the demands of each coalition of users, given the behavior of members external to the coalition. If the constructed linear program has a \emph{non-empty feasible set}, then the core of the game is nonempty and there exists a division of total utility such that no coalition has an incentive to deviate and cooperation is said to be \emph{stable}. If the feasible set of the linear program describing the core is empty, then the game exhibits oscillatory behavior among several configurations.

If the grand coalition (GC) has the highest utility among all possible configurations and the core is nonempty, then the GC is the stable outcome of cooperation. While CFGs have a unique definition of the core, PFGs do not have a unique notion of the core due to the dependence of the behavior of external coalitions. Assuming uniform behavior among external coalitions, several cores have been defined for PFGs to account for different behavior of coalitions \cite{Hafalir}. We now state the definitions of a few cores of PFGs suggested in the literature (note that this is not an exhaustive list that are relevant to the work herein).

The core of a game with \emph{rational expectations}, named the \emph{r-core}, is the feasible region for the set of linear inequalities:
\begin{eqnarray}
\sum_{i \in S} x_i & \geq & v_{\rho^{*}_S}(S; \{ S,\rho^{*}_S \} ), \forall S \subset \mathcal{K}, ~~ \sum_{i=1}^{K} x_i   =  v(\mathcal{K};\mathcal{K}), \nonumber \\
\rho^{*}_S  & = & \arg \max_{\rho_S} \sum_{G \in \rho} v(G; \{ S, \rho \}).
\label{e:rcore}
\end{eqnarray}
The \emph{r-core} models rational behavior among the remaining players \textit{i.e.,} all the other players excluding the deviating coalition try to maximize the sum utility, assuming that the deviating coalition cannot be changed anymore.

The core of the game with \emph{merging expectations}, named the \emph{m-core}, is the feasible region for the set of linear inequalities:
\begin{eqnarray}
\sum_{i \in S} x_i & \geq & v(S; \{ S,\mathcal{K} \setminus S \} ), \forall S \subset \mathcal{K}, ~~\sum_{i=1}^{K} x_i = v(\mathcal{K};\mathcal{K}).
\label{e:mcore}
\end{eqnarray}
Each coalition $S$ evaluates its utility by assuming that all the other players form a coalition irrespective of the actual partition of the users external to $S$.

The core of the game with \emph{cautious expectations}, named the \emph{c-core}, is the feasible region for the set of linear equalities:
\begin{eqnarray}
\sum_{i \in S} x_i & \geq & \min_{\rho} v(S;\{S,\rho\}), \forall S \subset \mathcal{K}, ~~ \sum_{i=1}^{K} x_i =  v(\mathcal{K};\mathcal{K}),
\label{e:ccore}
\end{eqnarray}
where the minimization is carried out over all partitions $\rho$ of the remaining users $\mathcal{K} \setminus S$. Let $\rho_{S}^{*}$ be the minimizing partition. Each coalition $S$ is \emph{guaranteed} a reward of $v(S;\{S,\rho_{S}^{*}$ \}), independent of the actual partition of the remaining users and the utility expected by each coalition is a conservative estimate of the actual obtainable utility.

Finally, the core with \emph{singleton expectations}, named the \emph{s-core}, as the feasible region of the set of inequalities:
\begin{eqnarray}
\sum_{i \in S} x_i & \geq & v(S; \{ S,(\mathcal{K} \setminus S) \}), \forall S \subset \mathcal{K}, ~~\sum_{i=1}^{K} x_i  =  v(\mathcal{K};\mathcal{K}),
\label{e:score}
\end{eqnarray}
where, $(\mathcal{K} \setminus S)$ denotes the partition containing all the singletons. This is in direct contrast to the \textit{m-core} where the utility of each coalition is computed by assuming that the rest of the users are in a single coalition.

\begin{figure}[ht]
\begin{minipage}[b]{0.5\linewidth}
\centering
\includegraphics[width=\textwidth]{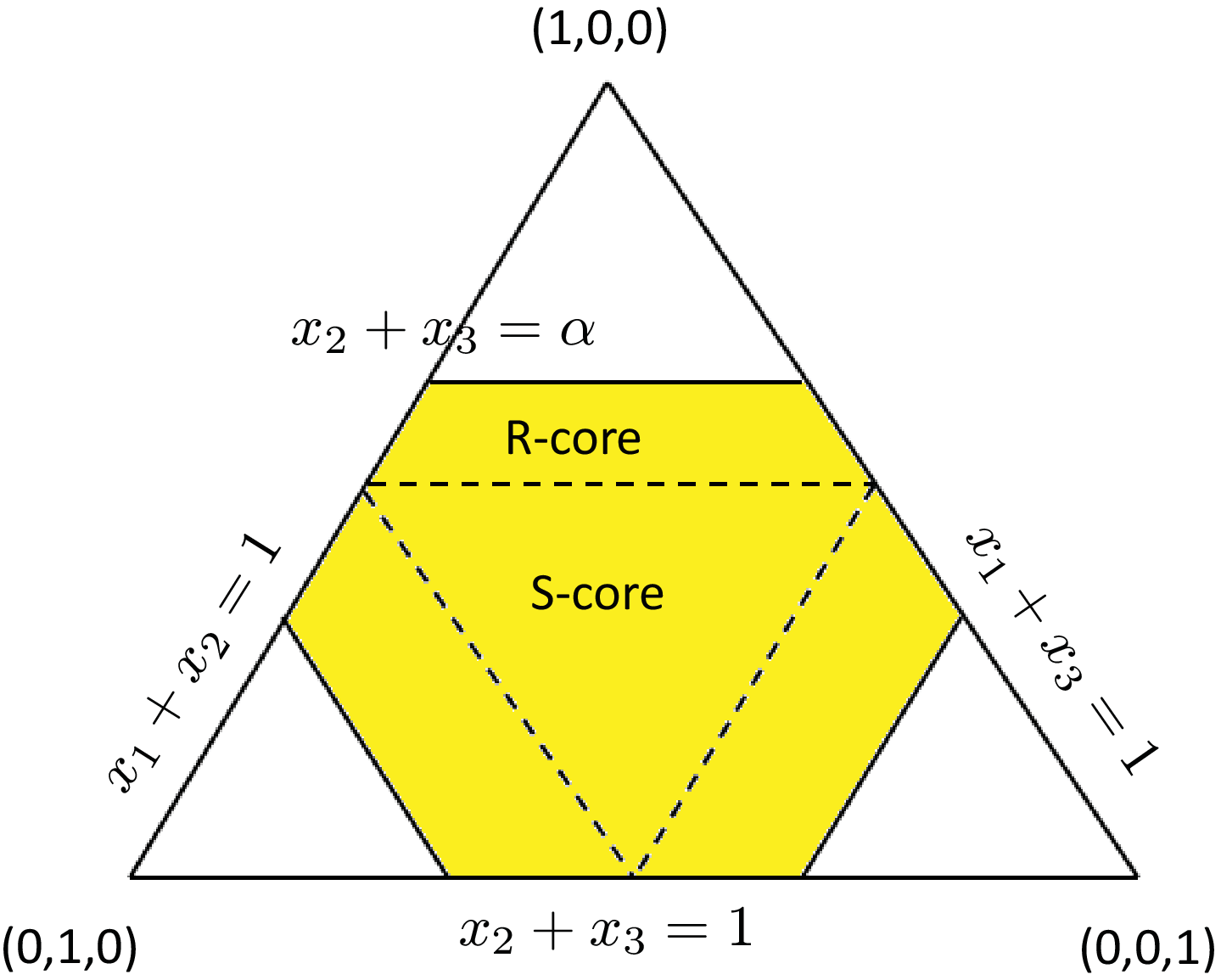}
\caption{Example of nonempty r-core (shaded region) and s-core (inside dotted region) for a symmetric scenario.}
\label{fig:core}
\end{minipage}
\hspace{0.5cm}
\begin{minipage}[b]{0.5\linewidth}
\centering
\includegraphics[width=0.8\textwidth]{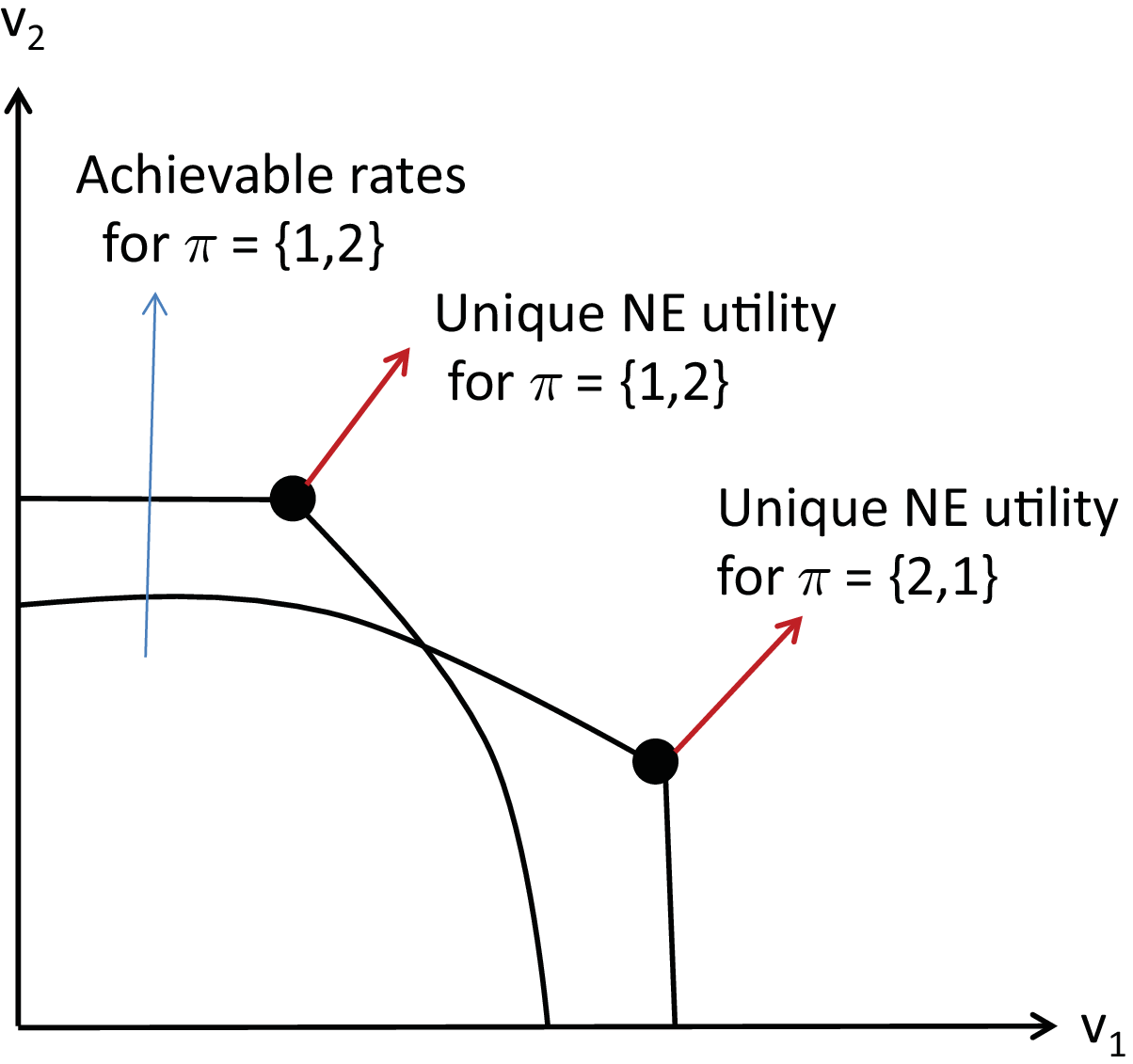}
\caption{Figure showing the NE rate points for an SIC single antenna receiver for different decoding orders.}
\label{fig:SIC_rate_point}
\end{minipage}
\end{figure}

\textit{Relationship between the various cores}: For PFGs with $r$-super-additivity, the rational behavior of external coalitions is to merge together to form the largest possible coalition and hence the \textit{r-core} is identical to the \textit{m-core} in which all the external players are treated as a single entity. For PFGs with negative externalities, the utility of a coalition is minimized when all the other coalitions in a partition operate in a unified manner and hence the \textit{c-core} is identical to the \text{m-core} in this scenario. In addition, for games with negative externalities, the constraints that define the \textit{s-core} are tighter than the constraints that define the \textit{m-core} and hence the \textit{s-core} is a subset of the \textit{m-core}. Fig. \ref{fig:core} shows an example of a nonempty r-core and s-core for a symmetric scenario with super-additivity and negative externalities. To summarize, for games with $r$-super-additivity and negative externalities, we have that \textit{s-core} $\subseteq$ \textit{m-core} = \textit{r-core} = \textit{c-core} and for games with $r$-super-additivity and mixed externalities, we have that the \textit{m-core} = \textit{r-core}. In this paper, we primarily focus on the \emph{r-core} and note that similar results can derived for the other PFG cores.

\subsection{Determining nonemptiness of a core}
We now state a necessary and sufficient to determine the nonemptiness of a core. For any $ S \subseteq \mathcal{K} $, denote by $\underline{1}_S \in \mathbb{R}^{K}$ the characteristic vector of $S$ given by $ (1_S)_i = 1$ when $i \in S $ and $(1_S)_i = 0$ otherwise. The collection $(\lambda_S), S \subseteq \mathcal{K}$ of numbers in $[0,1]$ is a \textit{balanced collection of weights} if for every player $k$, the sum of $(\lambda_S)$ over all the coalitions that contain the $kth$ player is 1, \textit{i.e.,} $ \sum_{S \subseteq \mathcal{K}} \lambda_S \underline{1}_S = \underline{1}_\mathcal{K} $. The Bondareva Shapley theorem \cite[p.262]{Osborne} states that a \textit{CFG} with transferable payoff has a nonempty core if and only if the game is balanced:
\begin{equation}
\sum_{S \subseteq \mathcal{K}} \lambda_S v(S) \leq v(\mathcal{K}) ~ \forall \lambda_S.
\label{e:bstheorem}
\end{equation}
Though the Bondareva Shapley theorem has been derived in the context of CFGs, it can be applied to PFGs to examine the nonemptiness of the various cores. It is well known that super-additive games may not satisfy the balancedness condition and can have empty cores \cite{Conitzer06}. Checking whether a game is balanced or not (\textit{i.e.,} verifying the nonemptiness of the core and thus the stability of cooperation) is in general a co-NP-complete problem \cite{Greco11}. In this paper, we exploit the structure of the utility function to verify the nonemptiness of the core in several regimes of interest.

\subsection{Games with empty cores}
\label{subsec:emptycore}
In several scenarios, it is feasible for a super-additive game to have an empty core wherein the GC has the highest utility, but cooperation is unstable. Such games exhibit oscillatory behavior as described in \cite{Mathur08}. Several approaches have been suggested in the literature to enforce stability of cooperation in such scenarios. We consider the \textit{m-core} for illustration. The $\epsilon$-core of a game is defined as the set of allocations such that
\begin{eqnarray}
\sum_{i \in S \subset \mathcal{K}} x_i & \geq & \min(v(S; \{ S,\mathcal{K} - S \}) - \epsilon,0), ~~\sum_{i=1}^{K} x_i =  v(\mathcal{K};\mathcal{K}).
\label{e:e-m-core}
\end{eqnarray}
In effect, the RX penalizes each coalition for leaving the GC. By choosing a large enough value of $\epsilon$, the $\epsilon$-core can always be made nonempty. The \textit{least core} of the game is defined as the $\epsilon$-core for the smallest value of $\epsilon$ that makes the core nonempty. The least core can be obtained by solving the optimization problem $\epsilon^{*} = \min \epsilon $ subject to (\ref{e:e-m-core}). If cooperation is not stable, the RX can penalize each deviating coalition to the extent determined by $\epsilon^{*}$ by refusing to decode signals sent at a higher rate and thus enforcing the stability of the GC. Finally, we note that the amount of penalty imposed on each coalition in (\ref{e:e-m-core}) is an illustration and in general, the penalty can vary depending on the coalition structure and the problem at hand.

\section{Signal Model}
\label{sec:signal_model}
Let us consider a MIMO MAC scenario with $K$ users simultaneously transmitting to a $M$-antenna receiver. Assuming that the $kth$ user has $n_k$ TX antennas, the link between the $kth$ TX and the RX is modeled by a deterministic channel matrix $G_k$ of dimension $M \times n_k$. The signal at the RX can be expressed as
\begin{equation}
\underline{Y}_{M \times 1} = \sum_{k=1}^{K} G_k \underline{X}_k + \underline{Z}_{M \times 1},
\end{equation}
where $ Z \sim \mathcal{N}(\underline{0},N_0I_{M \times M}) $ is the additive white Gaussian noise and $\underline{X}_k$ is the transmitted signal from the $kth$ user. Throughout this paper, we assume that the transmitted signals are drawn from a Gaussian codebook with power constraints.

\subsection{Signal Model with Coalitions and User Cooperation}
We illustrate the signal model with user cooperation for a given partition $T = \{S_1,S_2,...,S_N\}$ of users $\mathcal{K} = \{1,2,...,K\}$. Assume that the cooperating users organize themselves into coalitions, forming a partition $T$ of the set of users. All the users in a coalition $S_n$ act as a single virtual user and cooperate by jointly encoding the data to be transmitted to the receiver, thus acting as a virtual MIMO system. Note that only the users in each coalition cooperate with each other and users across different coalitions do not perform joint encoding. Let $H_n = [ G_{k_1}, G_{k_2}, \hdots, G_{k_{|S_n|}} ] $ be the effective channel matrix of dimension $M \times \sum_{j=1}^{|S_n|} n_j $ as seen by the $nth$ coalition. The signal at the RX can then be expressed as
\begin{equation}
\underline{Y}_{M \times 1} = \sum_{n=1}^{N} H_n \underline{X}_n + \underline{Z}_{M \times 1},
\label{e:rxsignal}
\end{equation}
where $[\underline{X}_n]_{\sum_{j=1}^{|S_n|} n_j \times 1 } $ is the transmitted vector for the coalition $S_n$ with a covariance matrix given by $Q_n = E [ \underline{X}_n \underline{X}^{H}_n ] $. We consider two types of power constraints in this paper : (1) a transmit sum power constraint for each coalition, \text{i.e.,} $\mbox{Tr}(Q_n) \leq \sum_{i=1}^{|S_n|} P_{S_n(i)} $ where $P_{S_n(i)}$ is the transmit sum power constraint of $ith$ user in the $nth$ coalition and (2) a per-antenna power constraint for each antenna of each user in a coalition, $ \mbox{diag}([Q_n]) \leq \mbox{diag}([P_{{S_n}(1),1}, P_{{S_n}(1),2},...]) $ where $P_{S_n(i),j}$ is the power constraint on the $jth$ antenna of the $ith$ user in coalition $S_n$. . We note that the results in this paper rely on key capacity computations that exist in literature for the above signaling models.  

\section{TX Cooperation Game with Successive Interference Cancelling Receiver}
\label{sec:tx_cooperation_intf}
Consider an arbitrary partition $T = \{S_1,S_2,...,S_N\}$ of the available TXs transmitting data to a common RX as described in Section \ref{sec:signal_model}. The RX first announces a decoding order $\pi$, a permutation of $(1,2,...,N)$, of the coalitions in $T$.  The TX cooperation game is analyzed in several stages: (1) Given $T$ and $\pi$, the interaction between the coalitions in $T$ is modeled as a non-cooperative game and the utilities that can be achieved in the configuration examined; (2) next, the properties which influence cooperative behavior among coalitions such as super-additivity, externalities \textit{etc.} are examined and (3) the stability of cooperation is analyzed by examining the core of PFGs.

\subsection{Non-cooperative game between TXs}
\begin{figure}
\centering{
    \includegraphics[width=7in,height=1.2in]{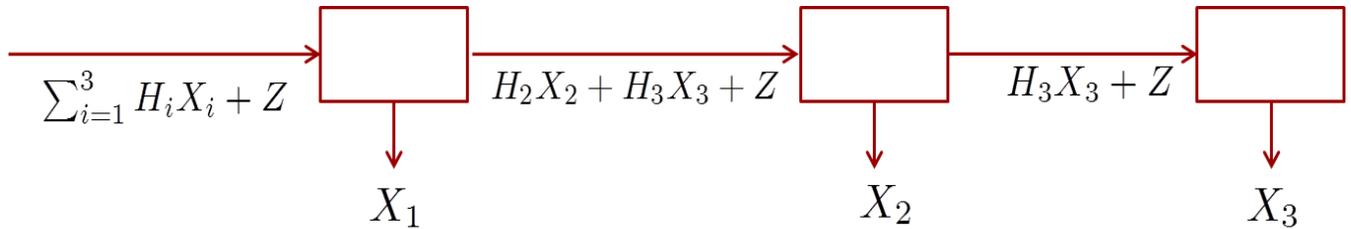}
\caption{Decoding in an SIC receiver}
\label{fig:SIC_receiver}}
\end{figure}

The RX signal is the sum of signals arriving from each coalition of TXs and is given by (\ref{e:rxsignal}). For a given partition $T$ and a SIC RX (see Fig. \ref{fig:SIC_receiver}), the strategy of the $kth$ coalition consists in choosing the vector of transmit covariance matrices $Q^{all}_{k} = \left( Q^{(1)}_k, Q^{(2)}_k, ... , Q^{(N!)}_k \right)$ each optimized for a given permutation $\pi$ of $\{1,2,...,N\}$. We begin by assuming a fixed decoding order and then extend our results to the case of time-sharing between decoding orders. Without loss of generality, let us assume that the decoding order of the coalitions is $\pi = (1,2,...,N)$. The utility obtained by each coalition is defined as the maximum achievable rate by all the TXs in the coalition, given the transmit covariance matrices of all other coalitions subject to the given power constraint. For a partition $T = \{S_1,...,S_N\}$, under the assumption of AWGN, the obtained utility by a coalition $S_n$ in $T$ adopting a transmit covariance matrix $Q_n$ can be expressed as
\begin{equation}
v(S_n;T) = \mbox{log} \left(  \frac{|N_0 I + H_n Q_n H^{H}_n  + \sum_{j=n+1}^{N} H_j Q_j^{*}H^{H}_j   |}{|N_0 I + \sum_{j=n+1}^{N} H_j Q_j^{*}H^{H}_j|}  \right).
\label{e:util_intf}
\end{equation}
Clearly, the utility obtained by each coalition $S_n$ depends on the structure of the partition $T$ and the strategies $Q_{-n}$ adopted by other coalitions in $T$. Each coalition chooses an action which is the best response to the actions of the users in the other coalitions and hence $v(S_n;T) = v_n(Q_n,Q_{-n})$ is defined as the \emph{NE utility} for the $nth$ player and the optimizing $Q_n$ is defined as the \emph{NE achieving strategy}. As the utility achieved by each coalition is dependent on the choices of actions of other players, the cooperative game between the TXs is a PFG. This is in contrast to the CFG model considered previously in \cite{Mathur08,La03} where the utility achieved by each coalition is independent of the actions of other players. As stated in Section \ref{sec:intro}, the PFG can capture the exact interference experienced by each coalition and thus is well suited to model scenarios arising in several wireless networks.

We now examine the existence of a NE of the non-cooperative game between the coalitions in $T$ which ensures that the NE achievable strategies and thus the NE utility always exist for the coalitions in a partition. The existence of a NE can be proved using the Kakutani fixed point theorem \cite{Osborne,Belmega10}. For illustration, we assume a per-antenna power constraint. Define the set $A_n = \{ Q_{n} | \mbox{diag}([Q_n]) \preceq \mbox{diag}([P_{{S_n}(1),1}, P_{{S_n}(1),2} , ... , ]) \} $, \textit{i.e.}, $A_n$ is the set of all covariance matrices which satisfy the per-antenna power constraint for the $nth$ coalition. Clearly, $A_n$ is a compact and convex set. The utility function $v(Q_n,Q_{-n})$ is continuous in $Q_i$ for all $i=1,2,...,N$ and is concave in $Q_n$ \cite{Kim06,Belmega10}. This satisfies all the conditions of the Kakutani fixed point theorem \cite{Osborne} and ensures the existence of a NE for the non-cooperative game between the coalitions in a partition.

\subsubsection{Uniqueness of NE} We now discuss the uniqueness of the NE achieving strategies and NE utilities. Note that the utility function $v(Q_n,Q_{-n})$ in (\ref{e:util_intf}) is concave, but not be strictly concave in $Q_n$ for a general case. This suggests that there \emph{may} exist many choices of $Q_n$ which result in the same value of the utility function. In the literature, uniqueness of NE has been used to suggest the uniqueness of NE achieving strategies which in turn implies the uniqueness of NE utilities. However, in this paper, we make the distinction between uniqueness of NE achieving strategies and NE utilities due to the non-strict concavity of $v_n(Q_n,Q_{-n})$ in (\ref{e:util_intf}).

\subsubsection{Relevance of uniqueness of NE utilities} As defined in Section \ref{sec:prelims}, the core of a PFG is a linear program which describes the demands of each coalition under an assumption on the behavior of external coalitions. The values of the utilities used in defining the PFG core in (\ref{e:rcore}), (\ref{e:mcore}), (\ref{e:ccore}) and (\ref{e:score}) are the NE utilities which are derived from the non-cooperative game between the coalitions in a given partition. Thus, we see that the uniqueness of the NE utilities allows the core to be well defined. If the NE utilities are not unique, then the core can be written for each combination of possible values of utilities. In the paper, we define the core as the union of all the cores that are obtained by choosing from the various possible values of the NE utilities.

The uniqueness of NE achieving strategies can be checked by deriving a sufficient condition for uniqueness. In \cite{Rosen65,Samson09,Belmega10} a sufficient condition called \emph{diagonally strict concavity} (DSC) has been derived to verify the uniqueness of NE achieving strategies. The DSC condition can be interpreted as the case where a player's utility function is more sensitive to the choice of his own actions as compared to the actions of all the other players. In this paper, we generalize the DSC condition proposed in \cite{Belmega10}. Though we derive this condition for the scenario with per-antenna power constraints, we note that the condition is applicable to the scenario with sum power constraints as well.

\begin{lemma}
If $ \left ( {\tilde Q}_1, {\tilde Q}_2,...,{\tilde Q}_N \right ) $ and $ \left ({\hat Q}_1, {\hat Q}_2,..., {\hat Q}_N \right) $ be two sets of covariance matrices which are NE to the non-cooperative game between the TXs in (\ref{e:util_intf}), then
\begin{equation}
C_n = \textrm{Tr} \left [ ({\tilde Q}_n  - {\hat Q}_n)( \nabla_{Q_n} v_n({\hat Q}_n,{\hat Q}_{-n}) - \nabla_{Q_n} v_n({\tilde Q}_n,{\tilde Q}_{-n})) \right ] \leq 0, \forall n=1,2,...,N.
\end{equation}
\label{lemma:DSC_lemma}
\end{lemma}
\begin{IEEEproof}
By the definition of an NE, the covariance matrices are the solutions to the optimization problem in
(\ref{e:PFG_utility_wo_intf}). The Lagrangian $\mathcal{L}_n$ for the maximization in (\ref{e:PFG_utility_wo_intf}) can be written as
\begin{eqnarray}
\mathcal{L}_n &  =  &  v_n(Q_n,Q_{-n}) + \mbox{Tr}(L_n Q_n) - \mbox{Tr} ( D_n (Q_n - R_n) ), 
\end{eqnarray}
where $L_n$ is a positive semi-definite matrix, $D_n = \mbox{diag}\left (\lambda_{S_n(1),1},...,\right)$ and $R_n = \mbox{diag}\left(P_{S_n(1),1},...,\right)$ are the diagonal matrices containing the Lagrange multiplier coefficients and the power constraint values respectively. For non-trivial power constraints, the Slater condition is satisfied and the Karush Kuhn Tucker (KKT) conditions can be written as
\begin{enumerate}[(a)]
\item $ {\tQ}_n \succeq 0,  {\hQ}_n \succeq 0 $.
\item $ \mbox{diag}([\tQ_n]) \preceq \mbox{diag}([P_{{S_n}(1),1}, P_{{S_n}(1),2},...]) $ and $ \mbox{diag}([\hQ_n]) \preceq \mbox{diag}([P_{{S_n}(1),1}, P_{{S_n}(1),2},...])$.
\item $ \mbox{Tr} ( {\tilde L}_n  {\tilde Q}_n ) = 0 $ and $\mbox{Tr} ( {\hat L}_n  {\hat Q}_n ) = 0 $.
\item $ \mbox{Tr} ( {\tilde D}_n ( {\tilde Q}_n - R_n  )) = 0  $ and $ \mbox{Tr} ( {\hat D}_n ( {\hat Q}_n - R_n  )) = 0 $.
\item $ \nabla_{Q_n} v({\tilde Q}_n,{\tilde Q}_{-n}) + {\tilde L}_n - {\tilde D}_n = 0 $ and $ \nabla_{Q_n} v({\hat Q}_n,{\hat Q}_{-n}) + {\hat L}_n - {\hat D}_n = 0 $.
\end{enumerate}
Now using the KKT conditions to evaluate and simplify $C_n$, we get
\begin{eqnarray}
-C_n & = &  \mbox{Tr} \left [ ( {\tilde Q}_n - {\hat Q}_n )
  \left ( \nabla_{Q_n} v({\tilde Q}_n,{\tilde Q}_{-n}) - \nabla_{Q_n} v({\hat Q}_n,{\hat Q}_{-n})  \right )     \right ]  \nonumber \\
  & \stackrel{(e)}{=} &  \mbox{Tr} \left [ ( {\tilde Q}_n - {\hat Q}_n ) \left ( ({\tilde D}_n - {\tilde L}_n) - ({\hat D}_n  - {\hat L}_n )  \right )      \right ]  \nonumber \\
  & = &  \mbox{Tr} \left [  {\tilde Q}_n {\tilde D}_n -  {\tilde Q}_n {\tilde L}_n  -  {\tilde Q}_n {\hat D}_n  +  {\tilde Q}_n {\hat L}_n - \left (  {\hat Q}_n {\tilde D}_n +  {\hat Q}_n {\tilde L}_n  +  {\hat Q}_n {\hat D}_n  - {\hat Q}_n {\hat L}_n  \right ) \right ] \nonumber \\
  & \stackrel{(c)}{=} &  \mbox{Tr} \left [  {R}_n {\tilde D}_n  -  {\tilde Q}_n {\hat D}_n  +  {\tilde Q}_n {\hat L}_n  - {\hat Q}_n {\tilde D}_n +  {\hat Q}_n {\tilde L}_n  +  {R}_n {\hat D}_n \right ] \nonumber \\
  & \stackrel{(a)}{\geq} & \mbox{Tr} \left [ {\tilde D}_n ( R_n - \tQ_n ) + {\hat D}_n (R_n - \hQ_n) \right ] \stackrel{(b,d)}{\geq} 0 \nonumber.
\end{eqnarray}
\end{IEEEproof}
The sequence of equalities and inequalities directly follow from the KKT conditions. Lemma \ref{lemma:DSC_lemma} shows that if there exist at least two equilibria of the non-cooperative game, then $ C_n \leq 0 ~ \forall n=1,2,...,N $.

\textit{Remark:} From Lemma \ref{lemma:DSC_lemma}, we can infer that if $C_n > 0$ for at least one value of $n$, then the NE achieving strategy is unique. This is a refinement of the condition in \cite{Rosen65,Belmega10} where one needed to show that $ C = \sum_{n=1}^{N} C_n > 0 $ for the NE achieving strategy to be unique. The DSC condition derived in \cite{Rosen65,Belmega10} holds for problems in which the strategy sets of each player are coupled with each other. In contrast, the refined DSC condition in our paper only holds for games where the strategy sets of each player are independent of each other and thus is restricted to the smaller class of NE problems.

\begin{lemma}
For two feasible strategies $ (\tQ_1,\tQ_2,...,\tQ_N) $ and $ (\hQ_1,\hQ_2,...,\hQ_N) $, $C = \sum_{n=1}^{N} C_n \geq 0 $.
\label{lemma:feasible_strategies_lemma}
\end{lemma}
\begin{IEEEproof}
Evaluating $C$ for any two feasible strategies $ (\tQ_1,\tQ_2,...,\tQ_N) $ and $ (\hQ_1,\hQ_2,...,\hQ_N) $, we get
\begin{align}
C & =  \sum_{n=1}^{N} \mbox{Tr} \left [ (\tQ_n - \hQ_n) ( \nabla_{Q_n} v_n(\hQ_n,\hQ_{-n})  - \nabla_{Q_n} v_n(\tQ_n,\tQ_{-n}) ) \right ] \nonumber \\
& = \sum_{n=1}^{N} \mbox{Tr} \left [ \left (H_n \tQ_n H_n^{H} - H_n\hQ_n H_n^{H} \right ) \left ( (N_0 I + \sum_{j=n}^{N} H_j \hQ_j H^{H}_j )^{-1} - (N_0 I + \sum_{j=n}^{N} H_j \tQ_j H^{H}_j  )^{-1} \right ) \right ] \nonumber \\
& = \sum_{n=1}^{N} \mbox{Tr} \left [ (A_n - B_n ) \left (  (\sum_{j=1}^{n} B_j )^{-1} - (\sum_{j=1}^{n} A_j )^{-1}     \right ) \right ] \geq 0.
\end{align}
where the matrices $A_n$ and $B_n$ are defined as $ A_1 = N_0 I + H_N \tQ_N H^{H}_N $, $ B_1 = N_0 I + H_N \hQ_N H^{H}_N $, $A_{N-n+1} = H_n \tQ_n H^{H}_n$ and $B_{N-n+1} = H_n \hQ_n H^{H}_n $ for $n \geq 2$ and the last inequality is true from \cite{Belmega10}.
\end{IEEEproof}

\begin{thm}
The \emph{NE utility} of the non-cooperative game between the TXs in a partition $T$ for a given decoding order $\pi$ is unique.
\label{thm:uniqueness_of_utility}
\end{thm}
\begin{IEEEproof}
From Lemma \ref{lemma:DSC_lemma} and Lemma \ref{lemma:feasible_strategies_lemma}, we infer that given any two NE achieving strategies of the game $ C_n = 0 ~\forall ~ n = 1,2,...,N $. Substituting for $C_n$, we get,
\begin{equation}
\mbox{Tr} \left [ \left (H_n \tQ_n H_n^{H} - H_n\hQ_n H_n^{H} \right ) \left ( (N_0 I + \sum_{j=n}^{N} H_j \hQ_j H^{H}_j )^{-1} - (N_0 I + \sum_{j=n}^{N} H_j \tQ_j H^{H}_j  )^{-1} \right ) \right ]  = 0.
\end{equation}
Using $C_N$ = 0 and the matrix trace inequality in \cite{Belmega10}, we get that $ H_N \tQ_N H^{H}_N = H_N \hQ_N H^{H}_N $. Now using $C_{N-1} = 0 $ and $ H_N \tQ_N H^{H}_N = H_N \hQ_N H^{H}_N $, we can show that $ H_{N-1} \tQ_{N-1} H^{H}_{N-1} = H_{N-1} \hQ_{N-1} H^{H}_{N-1} $. Continuing this approach, we can show that $ H_n \tQ_n H_n^{H} = H_n\hQ_n H_n^{H} ~ \forall ~ n = 1,2,...,N $. Now, substituting in the utility function in (\ref{e:util_intf}), it is clear that all NE achieving strategies have the same NE utility and hence the NE utility of the non-cooperative game between the TXs in a partition for a given decoding order is unique.
\end{IEEEproof}
We emphasize that Theorem \ref{thm:uniqueness_of_utility} only shows that the NE utility for all the NE achieving strategies is the same and hence unique. However, there may exist several achievable strategies which would achieve the NE utility. The uniqueness of the NE utility implies that, given the decoding order, each TX can \textit{unambiguously} evaluate the utility its obtains in a given partition.

\subsubsection{Evaluating the utility function}
For a coalition $S_n$ in partition $T$, the utility function as defined in (\ref{e:util_intf}) is the maximum achievable rate over the MIMO channel between the cooperating TXs and the RX, given the interference of all the other coalitions and the decoding order $\pi$. For the scenario with sum power constraints, the NE utility in (\ref{e:util_intf}) can be computed using sequential iterative water filling \cite{Yu04}. On the other hand, for a scenario with per-antenna power constraints, deriving a closed form expression for the capacity remains an open problem and \cite{Vu11MIMO} evaluates the capacity in terms of the the variables of the convex dual problem. However, a closed form solution can be derived for both the sum power and per-antenna power constraint scenario and a single antenna RX ($M=1$) and we state the NE utility function for this scenario.

Let us define $h_{{S_n}(i),j}$ as the channel gain from the $jth$ antenna of the $ith$ user in the $nth$ coalition to the RX. Using the capacity results in \cite{Yu04} and \cite{Vu11MIMO}, the NE utility for a coalition of TXs with a sum power constraint can be written as
\begin{equation}
v(S_m;T) = \mbox{log} \left ( \frac{ N_0 + \sum_{n=1}^{N} \left ( \sum_{i=1}^{|S_n|} \sum_{j}  |h_{{S_n}(i),j}|^2 \right )  \left ( \sum_{i=1}^{|S_n|} P_{S_n(i)} \right )  }{ N_0 + \sum_{n = m + 1}^{N} \left ( \sum_{i=1}^{|S_n|} \sum_{j} |h_{{S_n}(i),j}|^2 \right )  \left ( \sum_{i=1}^{|S_n} P_{S_n(i)} \right )  } \right ),
\label{e:util_single_antenna_SP}
\end{equation}
and with a per-antenna power constraint can be expressed as
\begin{equation}
v(S_m;T) = \mbox{log} \left ( \frac{ N_0 + \sum_{n=1}^{N} \left ( \sum_{i=1}^{|S_n|} \sum_{j}  |h_{{S_n}(i),j}| \sqrt{P_{{S_n}(i),j}}  \right )^2  }{  N_0 + \sum_{n = m + 1}^{N} \left ( \sum_{i=1}^{|S_n|} \sum_{j}  |h_{{S_n}(i),j}| \sqrt{P_{{S_n}(i),j}}  \right )^2 } \right ).
\label{e:util_single_antenna_PP}
\end{equation}
It is clear that beamforming achieves the NE utility for the single antenna RX in both scenarios. The key difference between the two scenarios is that for the per-antenna power constraint the beam weight has its phase matched to the channel coefficient, but the amplitude is independent of the channel and fixed based on the power constraint. Computing the NE utility in practice involves full knowledge of the channel gains and the power constraints of each user at all the players. The evaluated NE utility is used in negotiations to form new coalition structures, in determining the benefits gained by merging with other coalitions and determining the stability of cooperation.

The NE utility can be used to evaluate the total utility achievable by the current coalition and is used in negotiations to form new coalition structures, by quantifying the benefits gained by merging with other coalitions and determining the stability of cooperation as demonstrated later in this paper. Fig. \ref{fig:SIC_rate_point} shows the NE utility for a partition with $2$ coalitions and a single antenna receiver for both possible decoding orders. It can be inferred from \cite{Yu04} that the NE utility for the SIC receiver is on the Pareto-optimal boundary of the capacity region of the MAC and thus all the NE of this game are efficient.

\subsection{TX Cooperation - Properties}
\label{subsec:SIC_TX_cooperation}
Cooperation between TXs (coalitions of TXs) over the MAC channel with an SIC receiver has two benefits: (1) Cooperating coalitions signal jointly which can result in an improvement in the achievable sum rate; (2) The decoding order of the combined coalition improves relatively in comparison to the decoding order of its member coalitions resulting in a further improvement in the achievable rate for the cooperating coalitions. We illustrate this with an example. Consider a MAC scenario with $4$ TX coalitions $S_1, S_2, S_3, S_4$ specified in the order in which they are decoded. Assuming that coalitions $S_1$ and $S_3$ cooperate with each other, the receiver first decodes coalition $S_2$. Next $S_1$ and $S_3$, which signal jointly, are decoded followed by $S_4$. Clearly, $S_1$ benefits by moving later into the decoding order and both $S_1$ and $S_3$ benefit by signaling jointly. Joint signaling is achieved in practice by jointly generating the codebook using the knowledge of the joint density function and requires complete channel state information and power constraints of all the users. We formalize this intuition in the following propositions which describe the super-additive property and externalities for coalition formation.

\begin{prop}
The TX cooperation game with SIC processing at the receiver is \emph{$r$-super-additive} and \emph{cohesive}, \textit{i.e.,} for two partitions $T_1 = (S_1,S_2,S_3,...,S_r,S_{r+1},...,S_N) $ with decoding order $\pi_1 = (1,2,...,N)$ and $T_2 = (S_{a_1},...,S_{a_t}, S_{b_1} \cup S_{b_2} ... \cup S_{b_r}, S_{t+r+1}, S_{t+r+2}, ..., S_{N} ) $ with decoding order $\pi_2 = (a_1,a_2,...,a_t, \newline b_{\{12...r\}}, {t+r+1},...,N)$.
\label{prop:super-additive-SIC}
\begin{equation}
v(S_{b_1} \cup S_{b_2} ... \cup S_{b_r}; T_2) \geq \sum_{i=1}^{r} v(S_{b_i}; T_1),
\end{equation}
where all the utilities are as computed in (\ref{e:util_intf}). Note that $ (a_1,...,a_t,b_1,...,b_r) $ is a permutation of $(1,2,...,t+r)$ satisfying $a_1 < a_2 < ... < a_t$ and $ b_1 < b_2 < ... < b_r $.
\label{lemma:super_additive_intf}
\end{prop}
\begin{IEEEproof}
Let ${\tilde Q}$ and ${\hat Q}$ be the NE achieving covariance matrix tuples of $T_1$ and $T_2$ respectively. Then,
\begin{eqnarray}
v( S_{b_1} \cup \hdots \cup S_{b_r} ; T_2 ) & = & I ( X_{ {b_1} \cup \hdots \cup {b_r} } ; Y | X_{a_1} ,\hdots, X_{a_t} ) |_{\hat Q}
                \stackrel{(a)}{\geq} I ( X_{{b_1} \cup \hdots \cup {b_r} } ; Y | X_{a_1} ,\hdots, X_{a_t} ) |_{\tilde Q} \nonumber \\
                & \stackrel{(b)}{=} &  I ( X_{b_1} ; Y | X_{a_1} ,\hdots, X_{a_t} )|_{\tilde Q} +  I ( X_{b_2} ; Y | X_{a_1} ,\hdots, X_{a_t}, X_{b_1} )|_{\tilde Q} + \hdots \nonumber \\
                && + I ( X_{b_r} ; Y | X_{a_1} ,\hdots, X_{a_t}, X_{b_1}, X_{b_2},\hdots,X_{b_{r-1}} )|_{\tilde Q} \nonumber \\
                & \stackrel{(c)}{\geq} & \sum_{i=1}^{r} I ( X_{b_i} ; Y | X_1, X_2,\hdots, X_{b_i - 1} ) |_{\tilde Q} = \sum_{i=1}^{r} v( S_{b_i} ; T_1 ),
\end{eqnarray}
where the inequality (a) follows from the assumption of independent signaling among the cooperating coalitions and the definition of the NE, (b) and (c) follow from the chain rule of mutual information and the fact that $ (a_1,...,a_t,b_1,...,b_r) $ is a permutation of $(1,2,...,t+r)$ such that $a_1 < a_2 < ... < a_t, b_1 < b_2 < ... < b_r $. Hence the TX cooperation game with SIC is $r$-super-additive. Clearly, when all the coalitions cooperate, the TX cooperation game with SIC is cohesive, \textit{i.e.,} the GC has the highest sum utility and hence is the only feasible result of cooperation.
\end{IEEEproof}

\begin{prop}
The TX cooperation game with a \emph{single antenna SIC receiver} has \emph{negative} externalities.
\label{prop:negative-externalities-SIC-1}
\end{prop}
\vspace{-0.2in}
\begin{IEEEproof}
Using the notation in Proposition \ref{lemma:super_additive_intf}, we first see that $ v(S_{t+r+i} ; T_2 ) = v(S_{t+r+i};T_1), ~ \forall i = 1,2,\hdots,N-t-r$ as the NE utility of coalition $S_{t+r+i}$ only depends on the undecoded coalitions in their respective partitions, which are identical for $T_1$ and $T_2$ (see \ref{e:util_intf}). For the other coalitions, assuming a per-antenna power constraint, we have
\begin{eqnarray}
v(S_{a_n}; T_2) = \mbox{log} \left ( \frac{N_{int} + \alpha_n^2 + ( \sum_{i=1}^{r} \alpha_{t+i} )^2 }{N_{int} + ( \sum_{i=1}^{r} \alpha_{t+i} )^2} \right ) ~  \mbox{and} ~ v(S_{a_n}; T_1) = \mbox{log}\left  ( \frac{N_{int} + \alpha_n^2 +  \sum_{i=1}^{r} \alpha^2_{t+i} }{N_{int} +  \sum_{i=1}^{r} \alpha^2_{t+i}  } \right ), \nonumber
\end{eqnarray}
where $n=1,\hdots,t$, $\alpha_n = \sum_{i=1}^{|S_n|} \sum_{j} |h_{S_n(i),j}| \sqrt{P_{S_n(i),j}}  $ and $ N_{int} = N_0 + \sum_{i=n+1}^{t} \alpha^2_{i} + \sum_{i=t+r+1}^{N} \alpha^2_{i} $. From the above expressions, it can be clearly seen that $v(S_{a_n};T_2) \leq v(S_{a_n} ; T_1) $ and hence the TX cooperation game has negative externalities for a one antenna RX. Note that this result also holds for the sum power constraint scenario as well.
\end{IEEEproof}

\begin{prop}
The TX cooperation game with a \emph{multiple antenna SIC receiver} has \emph{mixed} externalities.
\label{prop:negative-externalities-SIC-2}
\end{prop}
\begin{IEEEproof}
We show the proposition by constructing examples which have both positive and negative externalities. Consider a 3-user scenario each with a single antenna TX transmitting to a 2-antenna RX. Let $\underline{h}_k$ be channel gain vector from the $kth$ TX to the RX and let $P_k \leq 1 $ be the per-antenna power constraint and let $T_1 = \{ \{1\},\{2\} , \{3\} \} $ and $ T_2 = \{ \{1,2\},\{3\} \} $ and let the higher numbered users are decoded first. The utility obtained by user $3$ under the partition $T_1$ and $T_2$  can be expressed as
\begin{equation}
v(S_3;T_1) = \mbox{log} ( |N_0 I + \sum_{i=1}^{3} \uh_i \uh^{H}_i | ) - \mbox{log} ( | N_0 I + \sum_{i=1}^{2} \uh_i \uh^{H}_i | ).
\label{e:neg_value_T1}
\end{equation}
and
\begin{equation}
v(S_3;T_2) = \mbox{log} \left ( \frac{|N_0 I + H_{12} Q_{12}^{*} H^{H}_{12} + \uh_3 \uh^{H}_3 |}{| N_0 I + H_{12} Q_{12}^{*} H^{H}_{12} |}  \right ),
\label{e:neg_value_T2}
\end{equation}
respectively with
\begin{equation}
Q^{*}_{12} = \arg \max_{Q_{12}} \mbox{log} \left (  \frac{|N_0 I + H_{12} Q_{12} H^{H}_{12} |}{|N_0 I|} \right ),
\end{equation}
It can be numerically observed that for some realizations of the channel gains $v(S_3;T_1) \leq v(S_3;T_2)$ and for other realizations $v(S_3;T_1) > v(S_3;T_2)$. For example, when $N_0 = 1$, $h_1 = [1.17119,-0.1941]$, $h_2 = [-2.1384,-0.8396]$ and $h_3 = [1.3546,-1.0722]$, we have that $v(S_3;T_1) = 0.8580 \leq v(S_3;T_2) = 1.0023 $ while for $h_1 = [-1.5771,0.5080]$, $h_2 = [0.2820,0.0335]$ and $h_3 = [-1.3337,1.1275]$ we have that $v(S_3;T_1) = 0.7593 \geq v(S_3;T_2) = 0.7462$. Hence the TX cooperation game with a multiple antenna SIC receiver has mixed externalities, in general.
\end{IEEEproof}
To summarize, we have shown that for merging coalitions for TXs, the NE utility achieved by the merging coalitions is at least as large as the sum of the NE utilities achieved by the individual coalitions. $r$-super-additivity of the PFG implies that the GC, the coalition of all the TXs signaling jointly, has the highest sum utility and hence is the only feasible outcome of cooperation. On the other hand, negative externalities for the single antenna RX imply that as when TXs merge to form larger coalitions, the rate achievable by every other TX reduces. This in turn induces further cooperative behavior as each rational TX tries to improve its utility further (as long as individual allocations increase). For the multi-antenna RX, mixed externalities imply that no general prediction can be made without knowing the specific channels gains and power constraints.

\subsection{TX Cooperation - Stability}
Previously, we have discussed the properties of the utility function when several coalitions merge to form a larger coalition. We now determine the \emph{stability} of TX cooperation by verifying the nonemptiness of the core. Note that verifying the satisfiability of the Bondareva-Shapley theorem is co-NP-complete even for super-additive games \cite{Greco11} and hence showing the nonemptiness of the core is a difficult problem in general. In this paper, we analyze the nonemptiness of the core in the high SNR (low noise) and high SNR (low noise) regime. We first derive an approximation for the capacity in the low SNR regime.

\begin{lemma}
In the low SNR regime, \textit{i.e.,} $N_0 \rightarrow \infty$, the capacity achieved by a player (here player 1) under the \emph{sum power} constraint can be approximated as
\begin{equation}
v(Q_1,Q_{-1}) = \max_{\mbox{Tr}(Q_1) \leq P_1 } \mbox{log} \left ( \frac{| N_0I + H_1 Q_1 H_1^{H} + K_{intf} |}{| N_0 I + K_{intf} |} \right ) \approx \frac{\sigma^2_{H_1} P_1}{N_0},
\end{equation}
where $\sigma_{H_1}$ is the maximum singular value of $H_1$ and $K_{intf} = \sum_{j=2}^{K} H_j Q^{*}_j H_j $ is the interference from all other users.
\label{lemma:lowSNRcapacity}
\end{lemma}
\begin{IEEEproof}
We begin by showing that at low SNR, the channel capacity is maximized by allocating all the power to the dominant eigen-mode of the channel. \begin{align}
v_1(Q_1,Q_{-1}) & = \max_{Q_1} \mbox{log} \left ( \frac{| N_0I + H_1 Q_1 H_1^{H} + K_{intf} |}{| N_0 I + K_{intf} |} \right ) = \max_{Q_1} \mbox{log} \left (  \frac{|N_0 I + {\tilde H}_1 Q_1 {\tilde H}^{H}_1 |}{|N_0 I|}  \right ), \nonumber \\
& \approx \max_{Q_1} \mbox{log} \left (  \frac{|N_0 I + U_1 \Sigma_1 V_1^{H} Q_1 V_1 \Sigma_1 U^{H}_1 |}{|N_0 I|}  \right ) = \max_{\mbox{Tr}(D_1) \leq P_1}  \left (  \frac{|N_0 I +  \Sigma_1 D_1  \Sigma_1  |}{|N_0 I|}  \right ) \nonumber \\
& = \max_{\mbox{Tr}(D_1) \leq P_1} \sum_{i} \mbox{log} \left ( 1 + \frac{\sigma^2_i d_i}{N_0} \right ) \stackrel{(a)}{=} \mbox{log} \left( 1 +  \frac{\sigma^2_{max} P_1}{N_0} \right ) \stackrel{(b)}{\approx} \frac{\sigma^2_{max} P_1}{N_0},
\end{align}
where $ {\tilde H}_1 = (I + \frac{1}{N_0} K_{intf})^{-1/2} H_1 $ and $ {\tilde H}_1 \approx H_1 $ as $N_0 \rightarrow \infty$, (a) is true as allocating all the power to the dominant eigen-mode maximizes the expression. Note that the capacity does not depend on any of the interferers in this regime and \emph{hence there are no externalities in low SNR regime}.
\end{IEEEproof}

\begin{thm}
In the low SNR regime, \textit{i.e.,} $ N_0 \rightarrow \infty$, the TX cooperation game with a \emph{sum power} constraint has a nonempty core.
\label{prop:nonempty_SIC_low_SNR}
\end{thm}
\begin{IEEEproof}
From Lemma \ref{lemma:lowSNRcapacity}, we know that the capacity at low SNR is effectively independent of the interference experienced by the coalition under consideration. Before showing the nonemptiness of the core, we derive the relation between the joint utilities of cooperating TXs and the utility of each individual TX. Consider two cooperating coalitions with channel gain matrices $H_1$ and $H_2$ respectively. The channel gain matrix of the combined coalition can be written as $H = [ H_1 | H_2 ] $. Using the fact that $H H^{H} = H_1 H^{H}_1 + H_2 H^{H}_2 $, we get that $ \sigma^2_{H_i} \leq \sigma^2_{H} \leq \sigma^2_{H_1 } + \sigma^2_{H_2} $ where $ \sigma_{H}$, $ \sigma_{H_1}$ and $\sigma_{H_2}$ are the maximum singular values of $HH^{H}$, $H_1 H_1^{H}$ and $H_2 H_2^{H}$ respectively.

Now assuming that there are $K$ TXs indexed by $\mathcal{K} = \{1,2,...,K\}$, the necessary and sufficient condition for the nonemptiness of the core is given by Bondareva-Shapley theorem from (\ref{e:bstheorem}):
\begin{equation}
\sum_{S \subset \mathcal{K}}\lambda_S v_{S} \leq v_{K} \Rightarrow \sum_{S \subset \mathcal{K}}\lambda_S \frac{ \sigma^2_{H_S}P_S}{N_0}   \leq \frac{ \sigma^2_{H_\mathcal{K}} P_{\mathcal{K}} }{N_0}
\label{e:SIC_core_sum_power_lowSNR}
\end{equation}
where $v_S = \sigma^2_{H_S}P_S/N_0 $ is the utility of coalition $S$ from Lemma \ref{lemma:lowSNRcapacity}, $ \sigma_{H_{S}} $ is the maximum singular value of the combined channel matrix of cooperating TXs $H_S$, $P_S = \sum_{i \in S} P_i $ and $\lambda_S$ is a balanced collection of weights. By substituting the bounds on the singular values of an augmented matrix in (\ref{e:SIC_core_sum_power_lowSNR}) and comparing coefficients on both sides, it is easy to see that the Bondareva-Shapley theorem holds and hence the core of the TX cooperation game with sum power constraints is nonempty at low SNR.
\end{IEEEproof}

\begin{thm}
In the high SNR regime, \textit{i.e.,} $ N_0 \rightarrow 0$, the TX cooperation game has an empty core for both the sum power and per-antenna power constraints.
\end{thm}
\begin{IEEEproof}
We give an example to show that the game has an empty core at high SNR. Consider a 4-user MAC (K=4) with one antenna at each TX and the RX. The power constraint on each user is $P_i \leq 1$ and the noise variance is $N_0 = 1$ with each user having an identical channel gain of $h_i = 1$. The receiver performs SIC by decoding the users in the order specified by a permutation $\pi$. Using the above parameters and computing the utilities as in (\ref{e:util_intf}), it can easily be verified that the \emph{r-core} of this game is empty. By symmetry, we infer that the \emph{r-core} is empty for all decoding orders.
\end{IEEEproof}

\begin{figure}[ht]
\begin{minipage}[b]{0.45\linewidth}
\centering
\includegraphics[width=\textwidth]{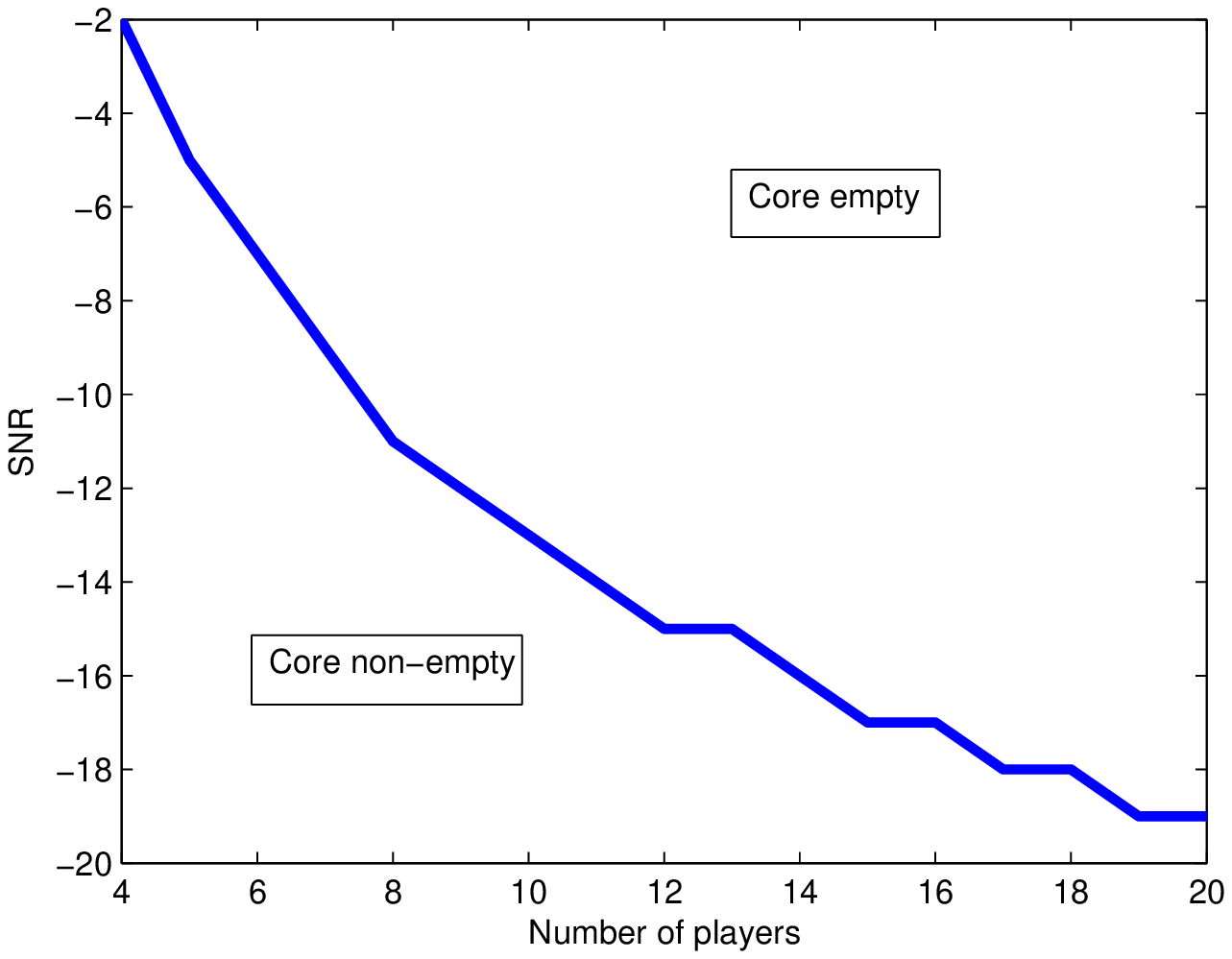}
\caption{Plot showing the boundary of the region between empty and nonempty core as a function of SNR and the number of players for a symmetric scenario and a single antenna SIC receiver with a fixed decoding order.}
\label{fig:SIC_feasibility_bound}
\end{minipage}
\hspace{0.5cm}
\begin{minipage}[b]{0.55\linewidth}
\centering
\includegraphics[width=0.8\textwidth]{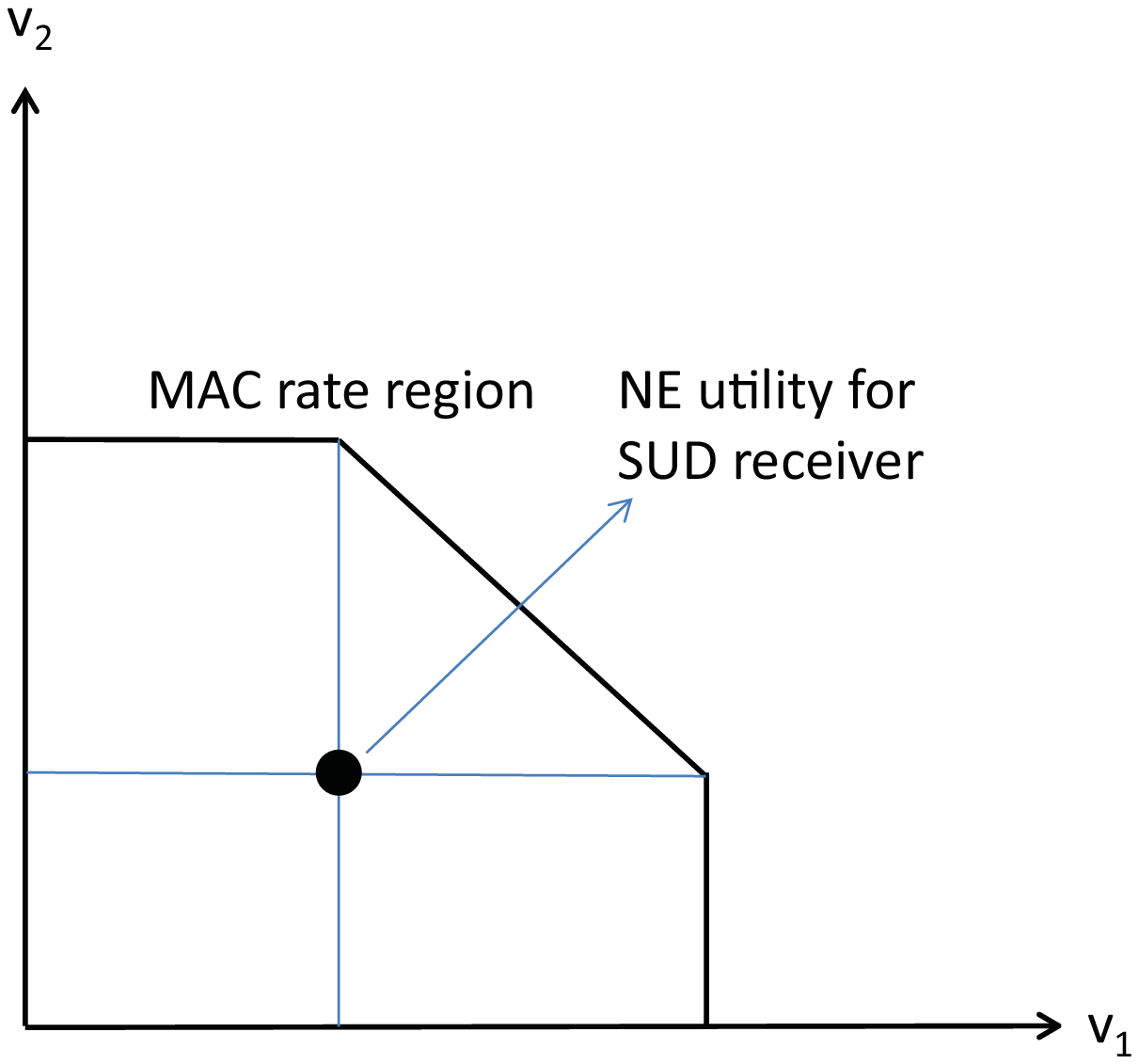}
\caption{Figure showing the NE rate point for an SUD single antenna receiver.}
\label{fig:SUD_rate_point}
\end{minipage}
\end{figure}

\emph{Discussion:} The optimal signaling strategy for each coalition at low SNR is beamforming along the best eigen-mode of the channel matched to the appropriate power constraint and the core is nonempty in this regime. This implies that the power gain due to beamforming is sufficient to make cooperation stable at low SNR. Note that as the NE utility at low SNR for a sum power constraint does not depend on the utility of other coalitions, the various cores of the PFG are identical in this regime and are all nonempty. For the case with per-antenna power constraints, we note that the nonemptiness of the core cannot be established for the low SNR case due to the lack of suitable approximation of capacity in this regime. In the high SNR regime, it is observed that the \emph{r-core} (similar statements can be made about other cores) is empty for all decoding orders in general. Fig. \ref{fig:SIC_feasibility_bound} shows a plot of the boundary between the regions of empty and nonempty \emph{r-core} as a function of the number of players and the SNR (Each player is assumed to have unit per antenna power constraint and unit channel gain and SNR is defined as $1/N_0$, \textit{i.e.,} the scenario where players are identical in all aspects other than the decoding order). Clearly, the core of the game is nonempty at low SNR and is empty at higher SNRs and the minimum SNR at which the core is nonempty reduces with the number of players. The empty core and hence instability of cooperation at high SNR can be attributed to the asymmetry between the TXs caused by a fixed decoding order. In contrast, noise dominates the interference in the low SNR regime and the decoding order becomes irrelevant. This removes the asymmetry between the players due to the decoding order and the core is nonempty in this regime. To enforce cooperation in the high SNR regime for a fixed decoding order, the RX can impose penalties on each coalition as described in (\ref{e:e-m-core}).

\subsection{SIC receiver with time sharing between decoding orders}
Previously, we have discussed the stability of TX cooperation for an SIC receiver assuming a fixed decoding order. We now consider the scenario in which time-sharing is permitted between the various decoding orders. From Theorem \ref{prop:nonempty_SIC_low_SNR}, the core for an SIC receiver is nonempty at low SNRs for a fixed decoding order and hence would be nonempty for a time shared SIC receiver. We now investigate the nonemptiness of the core at high SNRs.

Let $\Theta$ be the set of all probability distributions characterizing the time sharing of decoding orders. The set $ \Theta $ is a convex polyhedron whose corner points are distributions which assign probability $1$ to one of the decoding orders. Clearly all the probability distributions in $\Theta$ do not contribute to nonempty cores. Determining the subset of $\Theta$ for which the core is nonempty seems to be an intractable problem for now. Even the question of determining whether there exists a probability distribution of decoding orders for given channel gain matrices and power constraints appears to be a hard problem as verifying the Bondareva-Shapley theorem becomes highly nontrivial. To simplify the problem at hand, we consider an approximation to the utility function to understand the stability of cooperation. At high SNR, the dominant term can be considered a good approximation to the actual utility as the dominant term increases unboundedly while the other terms have a finite value. We now evaluate the \emph{r-core} of the game with this approximate utility function.

\begin{figure}[ht]
\begin{minipage}[b]{0.5\linewidth}
\centering
\includegraphics[width=3in,height=2.5in]{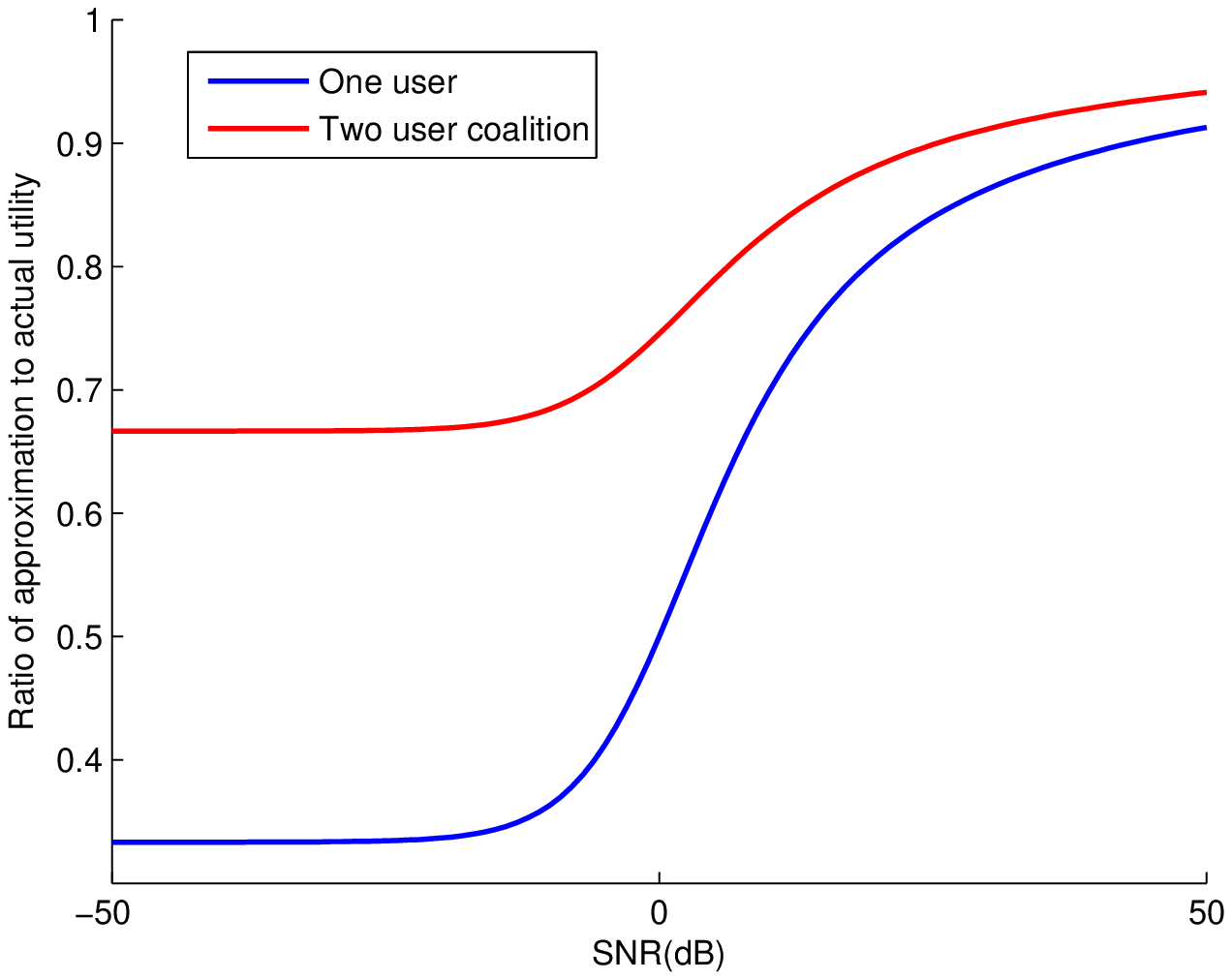}
\caption{Ratio of approximated to actual utility as a function of SNR for a 3-user MAC with single antenna TXs, unit channel gain and unit power constraint for a SIC RX. Note that the ratio approaches $1$ for very high SNR.}
\label{fig:ratio}
\end{minipage}
\hspace{0.5cm}
\begin{minipage}[b]{0.5\linewidth}
\centering
\includegraphics[width=3in,height=2.5in]{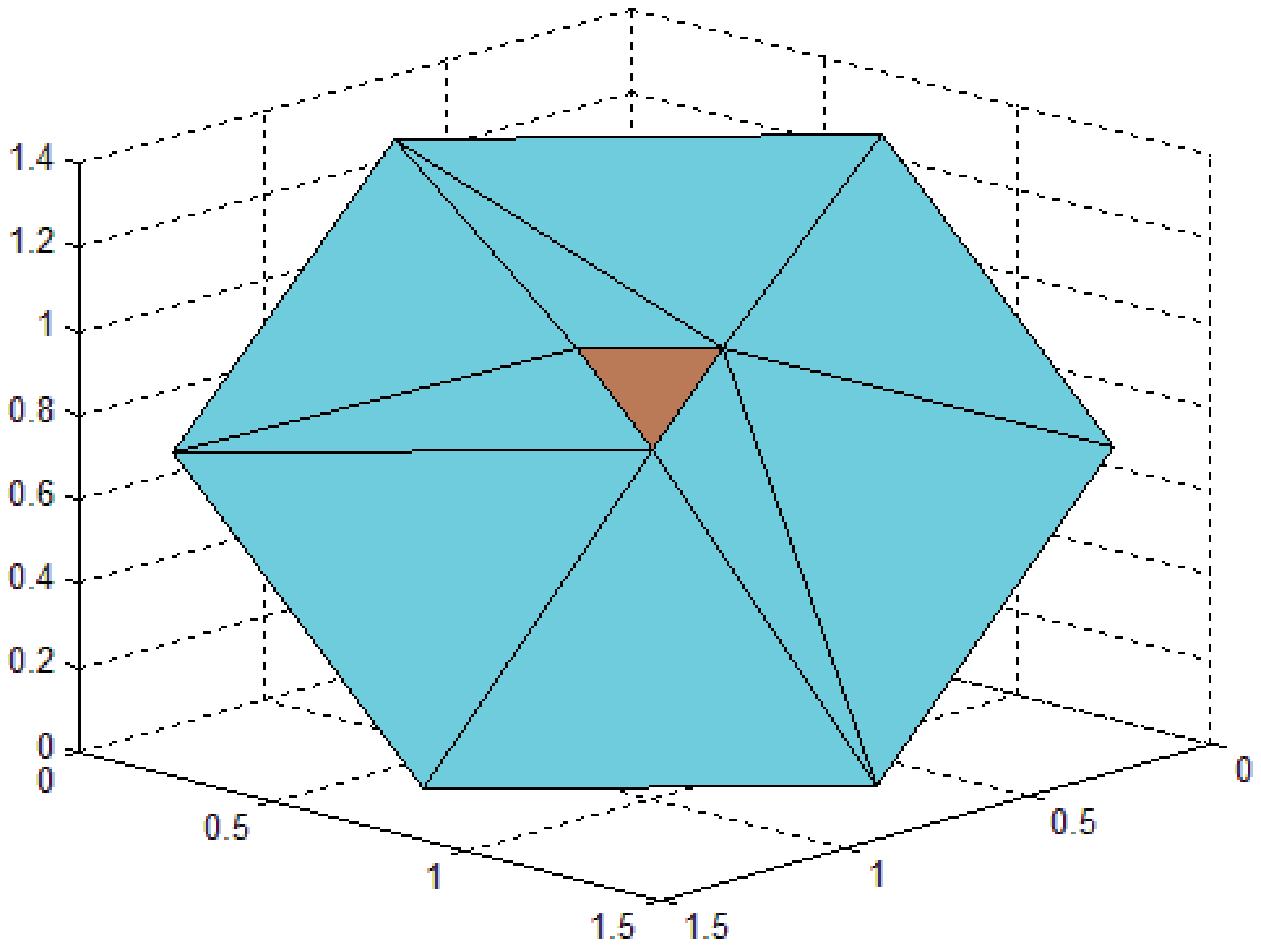}
\caption{Plot showing the r-core of the TX cooperation game for an SNR of 3dB for a 3-user symmetric MAC with single antenna TXs, unit channel gain and unit power constraint for an SIC RX with equal probability of time sharing between decoding orders using the exact utility function. The r-core is highlighted in red color.}
\label{fig:core_picture}
\end{minipage}
\end{figure}

\begin{thm}
The core of the TX cooperation game for an SIC receiver with equal probability of time sharing between all decoding orders is nonempty at high SNR to a first order approximation of utility.
\end{thm}
\begin{IEEEproof}
For the scenario in which all the decoding orders have equal probability, we first compute the utility function of a given coalition of players. We note that the dominating term in the utility function is the term in which a given coalition $S$ is decoded without any interference and increases unboundedly at high SNR. All the other terms are bounded at high SNR due to the interference present while decoding and can be ignored in this approximate analysis. The utility for coalition $S$ weighted with equal probability over all decoding orders can then be evaluated as
\begin{equation}
v_S = \frac{|S|}{K} \mbox{log} \left ( \frac{|N_0 I + H_S Q^{*}_S H^{H}_S |}{|N_0 I|}   \right ),
\end{equation}
where $|.|$ for sets denotes its cardinality. The first order approximation of utility can be interpreted as the scaled (by $|S|/K$) sum rate obtained when coalition $S$ signals with no interference. This also implies that the approximated utility does not depend on the strategies of other coalitions. Fig. \ref{fig:ratio} shows the ratio of the approximated utility to the actual utility for a 3-user symmetric MAC with unit channel gain and power constraints as a function of SNR. We see that in the high SNR regime, the ratio approaches unity, thus showing that the approximation is tight at high SNR. Substituting the utility in the Bondareva Shapley theorem in (\ref{e:bstheorem}), the necessary and sufficient condition to be satisfied can be expressed as
\begin{equation}
\sum_{S \subseteq \mathcal{K}} \lambda_S \frac{|S|}{K} \mbox{log} \left ( \frac{|N_0 I + H_S Q^{*}_S H^{H}_S |}{|N_0 I|}   \right ) \leq \mbox{log} \left ( \frac{|N_0 I + H_{\mathcal{K}} Q^{*}_{\mathcal{K}} H^{H}_{\mathcal{K}} |}{|N_0 I|}   \right ),
\label{e:main_core}
\end{equation}
for some balanced set of numbers $ \lambda_S$. The above condition is satisfied as the vector $\lambda_S \frac{|S|}{K} $ is a probability distribution and the utility for a coalition $S$ is always smaller than the utility of the GC. This ensures the core of the TX cooperation game with time sharing SIC receivers is nonempty at high SNR.
\end{IEEEproof}
This result implies that while fixing a decoding order results in an empty core at high SNR, time sharing between the decoding orders can result in an nonempty core (using an approximation to the utility function). Fig. \ref{fig:core_picture} shows the r-core of the TX cooperation game for a 3-user single antenna MAC with unit channel gain, unit power constraints, $N_0 = -3dB$ (SNR = 3dB) and time sharing with equal probability for all the decoding orders. The \emph{exact utility function} is used to describe the \emph{r-core} for this scenario. Clearly, the \emph{r-core} is nonempty showing the stability of TX cooperation in the high SNR regime as well (Note that the r-core is nonempty for SNRs more than 3dB also and the 3dB has been chosen for illustration in Fig. \ref{fig:core_picture}).

From a network design perspective, if a receiver implements SIC with time sharing when receiving data from several TXs, cooperation between the TXs by forming a virtual MIMO system and jointly signaling is stable and each transmitter improves his rate as compared to the non-cooperative scenario. As mentioned previously, accurate characterization of the nonemptiness of the core for high SNR using the exact utility function is still an open problem.

\section{TX Cooperation Game With Single User Decoding Receiver}
\label{sec:tx_cooperation}
In Section \ref{sec:tx_cooperation_intf}, we considered the problem of TX cooperation with SIC receivers and showed that TX cooperation is super-additive, has mixed externalities in general, cooperation is stable at low SNR for any decoding order and stable at high SNR when time sharing between decoding orders is permitted. In this section, we consider a RX which performs single user decoding (SUD) of the signals from each coalition. Following the methodology in Section \ref{sec:tx_cooperation_intf}, we investigate the stability of TX cooperation for such a receiver.

\subsection{Non cooperative game between TXs}

Consider an SUD receiver which decodes a given coalition's signal by treating all interfering signals from other coalitions as noise. The utility obtained by each coalition is defined as the maximum achievable rate by all the TXs in the coalition, given the transmit covariance matrices of all other coalitions subject to the given power constraint. For a partition $T = \{S_1,...,S_N\}$ of users, under the assumption of AWGN, the obtained utility by a coalition $S_n$ in $T$ adopting a transmit covariance matrix $Q_n$ can be expressed as
\begin{equation}
v(S_n;T) = v(Q_n,Q_{-n}) = \max_{Q_n} I(\underline{X}_n;\underline{Y})|_{Q_{-n}} = \max_{Q_n} \mbox{log} \left ( \frac{|N_0 I +  H_{n}  Q_{n} H^{H}_{n} + \sum_{j=1,j\neq n}^{N}  H_{j} Q_{j}  H^{H}_{j}  |}{|N_0 I + \sum_{j=1,j\neq n}^{N}  H_{j} Q_{j} H^{H}_{j} |} \right).
\label{e:PFG_utility_wo_intf}
\end{equation}
By definition, each coalition chooses an action which is the best response to the actions of the users in the other coalitions, hence is the cooperative game between the TXs is a PFG. This model of NE utility is in contrast to the approach adopted in \cite{Mathur08} where the SUD RX is designed assuming worst case interference from the other coalitions. We briefly illustrate the problem formulation and state the results from \cite{Mathur08} to highlight the key differences between the two approaches (Note that there is no similar CFG model for the TX cooperation game with an SIC RX).

\subsubsection{CFG Model}
In the jamming utility model in \cite{Mathur08}, the utility $u(S)$ of a coalition $S$ is defined as the maximum obtainable rate assuming worst case interference from TXs in $ S^{c} = \mathcal{K} \setminus S$.
\begin{equation}
u(S) = \min_{Q_{S^c}} \max_{Q_{S}}  I(X_{S}; Y) = \min_{Q_{S^c}} \max_{Q_{S}} \mbox{log} \left (\frac{|N_0 I + H_{S} Q_{S} H^{H}_{S} + H_{S^c} Q_{S^c} H^{H}_{S^c} |}{|N_0 I + H_{S^c} Q_{S^c} H^{H}_{S^c}|} \right),
\label{e:CFG_utility}
\end{equation}
where $Q_S$ and $Q_{S^{c}}$ are the transmit covariance matrices of the coalition $S$ and $S^{c}$ respectively and are constrained to satisfy the given power constraint. Clearly, $u(S)$ is a very conservative lower bound on the utility that can be obtained in practice as typical TXs will not attempt to jam the transmissions from $S$ at the cost of their own rate. Under the utility model in (\ref{e:CFG_utility}), \cite{Mathur08} shows that TX cooperation is cohesive. However, using a counter example, \cite{Mathur08} demonstrates an empty core in general and conjectures that the core is nonempty when all TXs roughly have similar channel gains. Thus, TX cooperation exhibits oscillatory behavior under this utility model. In contrast, we will demonstrate in this section that under the PFG model for utility, TX cooperation is stable in the high SNR and the low SNR regime.

\subsubsection{PFG model}
We now analyze the properties of the non-cooperative game between the coalitions in a partition of TXs to determine the NE utility of a given coalition. As in Section \ref{sec:tx_cooperation_intf}, the feasible sets of each user are convex and compact and utility function $v_n(Q_n,Q_{-n})$ is concave in $Q_n$ and continuous in $Q_i, i=1,2,...,N$. The existence of an NE then follows from the Kakutani fixed point theorem. We now examine the uniqueness of \emph{NE utility} for the MIMO SUD receiver.

\begin{lemma}
For two feasible strategies $ (\tQ_1,\tQ_2,...,\tQ_N) $ and $ (\hQ_1,\hQ_2,...,\hQ_N) $, we get $C = \sum_{n=1}^{N} C_n \geq 0 $, where $C_n$ is defined as in Lemma \ref{lemma:DSC_lemma}.
\label{lemma:feasible_strategies_lemma2}
\end{lemma}
\begin{IEEEproof}
Evaluating $ C $  for any given feasible strategies, we get
\begin{eqnarray}
C & = & \sum_{n=1}^{N} \mbox{Tr} \left [   (  \tQ_n - \hQ_n ) ( \nabla_{Q_n} v_n(\hQ_n,\hQ_{-n}) - \nabla_{Q_n} v_n( \tQ_n,\tQ_{-n}) )   \right ] \nonumber \\
  & = &  \mbox{Tr} \left [ \Bigg (\sum_{n=1}^{N} H_n ( \tQ_n - \hQ_n ) H^{H}_n \Bigg )  \Bigg (  \Bigg(N_0I + \sum_{n=1}^{N} H_n \hQ_n H^{H}_n \Bigg)^{-1} - \Bigg(N_0I + \sum_{n=1}^{N} H_{n} \tQ_n H^{H}_{n} \Bigg)^{-1} \Bigg ) \right ] \geq 0 \nonumber.
\end{eqnarray}
where the inequality follows from that the fact that for positive definite matrices $\tB = N_0I + \sum_{n=1}^{N} H_{n} \tQ_n H^{H}_{n} $  and $\hB = N_0I + \sum_{n=1}^{N} H_{n} \hQ_n H^{H}_{n}$ we have $ \mbox{Tr} \left \{ ( \tB - \hB ) (\hB^{-1} - \tB^{-1}) \right \} \geq 0 $ with equality only when $ \tB = \hB $.
\end{IEEEproof}
\begin{prop}
From Lemma \ref{lemma:DSC_lemma} and Lemma \ref{lemma:feasible_strategies_lemma2}, we can infer that if $ (\tQ_1,\tQ_2,...,\tQ_N) $ and $ (\hQ_1,\hQ_2,...,\hQ_N) $ are two NE achieving strategies of the game between the coalitions of TXs, then $\sum_{n=1}^{N} H_n \hQ_n H^{H}_n = \sum_{n=1}^{N} H_n \tQ_n H^{H}_n$.
\label{prop:uniquness_SUD}
\end{prop}
\begin{IEEEproof}
\begin{eqnarray}
C_n & = & \mbox{Tr} \left [ ({\tilde Q}_n  - {\hat Q}_n)( \nabla_{Q_n} v_n({\hat Q}_n,{\hat Q}_{-n}) - \nabla_{Q_n} v_n({\tilde Q}_n,{\tilde Q}_{-n})) \right ] \nonumber \\
    & = & \mbox{Tr} \left [ (  H_n ( \tQ_n - \hQ_n ) H^{H}_n )  \Bigg (  \Bigg(N_0I + \sum_{n=1}^{N} H_n \hQ_n H^{H}_n \Bigg)^{-1} - \Bigg(N_0I + \sum_{n=1}^{N} H_{n} \tQ_n H^{H}_{n} \Bigg)^{-1} \Bigg ) \right ] = 0 \nonumber \\
    & \Leftrightarrow & H_n \tQ_n H_n^{H} = H_n \hQ_n H_n^{H} ~ \forall n = 1,2,...,N (or) \sum_{n=1}^{N} H_n \hQ_n H^{H}_n = \sum_{n=1}^{N} H_n \tQ_n H^{H}_n \nonumber \\
    & \rightarrow & \sum_{n=1}^{N} H_n \hQ_n H^{H}_n = \sum_{n=1}^{N} H_n \tQ_n H^{H}_n.
\end{eqnarray}
This implies that if there exist two distinct NE achieving strategies for the non-cooperative game between the TXs in different coalitions, then the NE achieving strategies satisfy the condition $\sum_{n=1}^{N} H_n \hQ_n H^{H}_n = \sum_{n=1}^{N} H_n \tQ_n H^{H}_n$.
\end{IEEEproof}

\emph{Discussion:} While the uniqueness of NE utility can be demonstrated for the scenario with an SIC receiver, Proposition \ref{prop:uniquness_SUD} shows that this might not be true in general for an SUD receiver. However, numerical simulations suggest that the NE utility for this non-cooperative game is unique in general and a counter-example to this scenario has not been found. In \cite{Mertikopoulos11}, the uniqueness of NE of parallel multiple access channels has been investigated for a SUD receiver and it was shown that the NE achieving strategies and thus the NE utility for this class of channels is unique (almost surely). A special case of our scenario, wherein all the channel matrices $H_n$ are diagonal matrices can be modeled as a parallel multiple access channel and we can infer from \cite{Mertikopoulos11} that the NE utility in this scenario is unique (almost surely).

Consider a scenario in which each TX transmits to a common RX over a multipath fading channel. Orthogonal Frequency Division Multiplexing (OFDM) has been one of the schemes used in the literature to overcome multipath effects of the channel. OFDM transforms a multiple access channel into several parallel flat fading channels and the total power available at each user can be distributed among the subcarriers. This signaling scenario can be reinterpreted as a parallel MAC with a sum power constraint on the total power for all the subcarriers of a user. Using the result in \cite{Mertikopoulos11} clearly shows that uniqueness of NE utility for an OFDM signaling scenario when using an SUD receiver, thus demonstrating an important set of channels where the NE is indeed unique. However, the proof technique in \cite{Mertikopoulos11} does not directly generalize to the case of non-diagonal $H_n$ and the general problem of uniqueness for a MIMO SUD receiver remains an open problem.

Fig. \ref{fig:SUD_rate_point} shows one feasible NE utility for a scenario with 2 users and a single antenna RX. Note that this point is in the interior of the rate region as all the signals are decoded with interference.

\subsection{TX cooperation - Properties and Stability}
In Section \ref{subsec:SIC_TX_cooperation}, we demonstrated the properties of TX cooperation for an MIMO MAC with an SIC RX. We now state the properties of TX cooperation for the SUD RX.

\begin{prop}
The TX cooperation game is \emph{$r$-super-additive} and \emph{cohesive} (see (\ref{e:r_super_additive})), \textit{i.e.,} for two partitions $T_1 = (S_1,S_2,...,S_r,S_{r+1},...,S_N) $ and $ T_2 = (S_1 \cup S_2 \cup ... \cup S_r, S_{r+1}, S_{r+2},...,S_N) $ of $\mathcal{K}$,
$ v(S_1 \cup ... \cup S_r; T_2) \geq  \sum_{t=1}^{r} v(S_t ; T_1 ) $ where all the utilities have been computed as in (\ref{e:PFG_utility_wo_intf}).
\label{prop:super-additive-SUD}
\end{prop}

\begin{prop}
The TX cooperation game with a \emph{single antenna receiver} has \emph{negative} externalities, \textit{i.e.,} for two partitions $T_1 = (S_1, S_2, ... ,  S_r, S_{r+1} , ... , S_N) $ and $ T_2 = (S_1 \cup S_2 \cup ... \cup S_r, S_{r+1}, S_{r+2},...,S_N) $ of $\mathcal{K}$, $v(S_{r+i},T_2) \leq v(S_{r+i},T_1)$ for every $i=1,...,N-r$ and $\forall ~ r$.
\label{prop:negative-externalities-SUD-1}
\end{prop}

\begin{prop}
The TX cooperation game with a \emph{multiple-antenna receiver} has \emph{mixed} externalities, \textit{i.e.,} some realizations of the game has positive externalities and other realizations have negative externalities.
\label{prop:negative-externalities-SUD-2}
\end{prop}
The proof of Propositions \ref{prop:super-additive-SUD}, \ref{prop:negative-externalities-SUD-1} and  \ref{prop:negative-externalities-SUD-2} follows exactly along the lines of Propositions \ref{prop:super-additive-SIC}, \ref{prop:negative-externalities-SIC-1} and \ref{prop:negative-externalities-SIC-2} respectively in Section \ref{subsec:SIC_TX_cooperation} and are omitted here due to space limitations. We observe that the TX cooperation game for an SUD RX has the same properties as that of an SIC RX. This implies that when coalitions cooperate, the total utility achieved by the combined coalition is larger than the individual utilities and the grand coalition is the only feasible outcome of cooperation.

We now investigate the stability of the grand coalition for a SUD RX following the approach in Section \ref{sec:tx_cooperation_intf}. The key difference between the SUD and the SIC RXs is that the NE utility achieved for an SUD RX is not proven to be unique and hence there may exist multiple feasible NE utilities for a given partition of TXs. To address the issue of uniqueness, we define the core of a game to be union of the cores obtained by choosing all feasible combinations of NE utility. Thus, having a nonempty core from one combination of utilities ensures that the core of the game itself is nonempty ensuring the stability of cooperation. We now determine the stability of TX cooperation by verifying that the Bondareva Shapley theorem in Section \ref{sec:prelims} is satisfied.

\begin{thm}
In the high SNR regime, \textit{i.e.} $ N_0 \rightarrow 0$, the TX cooperation game with a SUD receiver has a nonempty core for both sum power and per-antenna power constraints.
\label{prop:nonempty_SUD_high_SNR}
\end{thm}
\begin{IEEEproof}
Let us first consider the \emph{r-core} of the game. The utility of coalition $S$ in a partition $T = \{S,S_1,...,S_N\} $ can be expressed from (\ref{e:PFG_utility_wo_intf}) as
\begin{equation}
v(S;T) =  \mbox{log} \left ( \frac{ | N_0 I + H_S Q^{*}_S H^{H}_S + \sum_{j=1}^{N} H_j Q^{*}_j H^{H}_j |}
    {| N_0 I + \sum_{j=1}^{N} H_j Q^{*}_j H^{H}_j | } \right ),
\end{equation}
where $(Q_S^{*},Q_1^{*},...,Q_{N}^{*})$ is an NE achievable strategy. We now show that for any balanced collection of weights $\lambda_S$, the Bondareva Shapley theorem in Section \ref{sec:prelims} holds at high SNR. Substituting for $v(S;T)$ in (\ref{e:bstheorem}), we get
\begin{eqnarray}
\sum_{S \subseteq N} \lambda_S v(S;T) & = & \sum_{S \subseteq N} \lambda_S
    \mbox{log} \left ( \frac{ | N_0 I + H_S Q^{*}_S H^{H}_S + \sum_{j=1}^{N} H_j Q^{*}_j H^{H}_j |}
    {| N_0 I + \sum_{j=1}^{N} H_j Q^{*}_j H^{H}_j | } \right ) \nonumber \\
    & \stackrel{(a)}{\leq} & \mbox{log} \left ( \frac{  |N_0 I + H_\mathcal{K} Q^{*}_\mathcal{K} H^{H}_\mathcal{K} |}{|N_0 I|} \right) = v(\mathcal{K};\mathcal{K}),
\end{eqnarray}
where $(a)$ is true for all partitions $T$ as the summation of the LHS has a finite value for large SNR and the RHS terms increases in an unbounded fashion, satisfying the conditions of the Bondareva Shapley theorem. This proves that the \emph{r-core} is nonempty at high SNR. Using similar arguments, it can be shown that all the various cores of a PFG defined in Section \ref{sec:prelims} are nonempty at high SNR showing that TX cooperation is stable at high SNR for an SUD receiver. In addition, we note that the proof is independent of the nature of the power constraints and hence is valid for scenarios with both sum power and per-antenna power constraints.
\end{IEEEproof}

\begin{thm}
In the low SNR regime, \textit{i.e.} $ N_0 \rightarrow \infty$, the TX cooperation game with a \emph{sum power} constraint and a SUD receiver has a nonempty core.
\label{prop:nonempty_SUD_low_SNR}
\end{thm}
\begin{IEEEproof}
We have observed in Theorem \ref{prop:nonempty_SIC_low_SNR} that the utility function at low SNR does not depend on the actual interference experienced. Using the same methodology as in Theorem \ref{prop:nonempty_SIC_low_SNR}, we can show that the TX cooperation game has a nonempty core and hence cooperation is stable at low SNR.
\end{IEEEproof}
As in the previous section, the nonemptiness of the core at low SNR for per-antenna power constraints cannot be determined analytically due to the lack of a suitable approximation to the capacity of a MIMO channel at low SNR. However, numerical simulations show that the core is indeed nonempty for this scenario as well. In summary, we note that ideal cooperation is stable for both the high SNR and low SNR regime for a SUD RX.

\subsection{Discussion}
\subsubsection{Which RX is better for enforcing cooperation ?}
In a MIMO MAC, this paper shows that for the low SNR regime, the GC is always a stable outcome of TX cooperation, under the assumption of no cooperation costs for both the SIC and SUD RXs. In this regime, all the power is allocated to one of the eigen modes of the channel and the power/beamforming gain that is obtained when multiple coalitions merge to form the GC is sufficient to ensure the stability of cooperation. On the other hand, the behavior in the high SNR regime is dominated by the effect of interference. For a SUD RX, any coalition deviating from the GC will experience significant interference and thus looses heavily by deviating from the GC. This ensures the stability of cooperation at high SNR as shown in Theorem \ref{prop:nonempty_SUD_high_SNR}. Numerical simulations suggest that this holds for all SNRs of interest in general though it can be analytically shown only for the regimes of extreme SNR. For a SIC RX and a fixed decoding order, we have shown that GC is not stable at high SNR and stable (with an approximation to the utility) for time shared decoding orders. To enforce cooperation at high SNR for a fixed decoding order, we can impose a penalty on any deviating coalitions as indicated in Section \ref{subsec:emptycore}. As the utility of the GC is the same irrespective of the RX used to decode the signals (it is the capacity of the virtual MIMO system formed when all the TXs cooperate), our analysis suggests that using a simpler SUD RX is better at enforcing cooperation in comparison to a more capable SIC RX.

\subsubsection{Fairness of rate allocation}
The core of the TX cooperation game, when nonempty, is a large convex set as it the feasible region of a set of linear inequalities. This implies that, in general, there exist uncountable number of divisions of utility such that no coalition of users has an incentive to deviate from the GC. While the core itself does not take into account fairness of allocation, a suitable fairness metric optimized over the core can result in a stable allocation which is fair to the extent allowed by the elements of the core.

\subsubsection{Information and computational requirements}
As mentioned previously, computing the NE utility for each configuration and then determining the nonemptiness of the core of a game, in general requires complete channel state information at all the players. For a $K$-player game, we see that $O(2^{K})$ utilities must be evaluated to express the core and evaluating nonemptiness of the core can be evaluated by solving a linear program with exponential ($2^{K} - 1$) number of constraints (see \cite{Greco11} for further details). While no cooperation costs are assumed in this analysis, sharing the channel state and codebook information in practice incurs costs and may reduce the NE utility that a coalition can achieve. In scenarios such as base station cooperation (see Fig. \ref{fig:network}), the backbone network connecting the base stations can be used to share the channel state and codebook information. In \cite{Saad09-TWC}, the cost of cooperation is modeled as a reduction in the power available for data transmission and \cite{Mathur08} considers a partial-decode and forward scheme to share the messages to be jointed transmitted. However, accurate modeling of the cost of cooperation is still an open problem in the literature.

\section{Conclusions}
\label{sec:conclusions}
The question of feasibility of cooperation between \emph{rational} nodes in a MIMO multiple access channel and whether there exists a fair division of the benefits of cooperation is addressed using partition form cooperative game theory to accurately model the effects of interference. The stability of the grand coalition, the coalition of all transmitters, for SUD and SIC receivers is examined. For an SUD receiver, TX cooperation is shown to be stable at high and low SNRs analytically and at all SNRs numerically, while for an SIC receiver with a fixed decoding order, TX cooperation is only stable at low SNRs where interference is negligible. However, using a high SNR approximation to the utility function, TX cooperation is stable with an SIC receiver implementing equal time sharing between decoding orders. In summary, our work demonstrates that under the assumption of zero costs, voluntary cooperation is feasible and stable between users in a MAC and every user benefits from cooperation.

\bibliographystyle{IEEEtran}
\bibliography{references}

\end{document}